\documentclass[twocolumn,aps,prd,floatfix,superscriptaddress,nofootinbib]{revtex4-2}

\usepackage{hyperref}
\usepackage{makecell}
\usepackage{tabularx}
\newlength\C
\usepackage{array}
\usepackage{latexsym, amsmath, amssymb,amsthm,amsopn,amsfonts, graphicx, epstopdf, multirow}
\usepackage{graphicx}
\usepackage{color}
\usepackage{amssymb}
\usepackage{amsmath}
\usepackage{natbib}

\usepackage{textcomp}
\usepackage{hyperref}
\usepackage{placeins}
\usepackage[dvipsnames]{xcolor}
\newcolumntype{P}[1]{>{\centering\arraybackslash}p{#1}}

\def\simlt{\lower.5ex\hbox{$\; \buildrel < \over \sim \;$}}
\def\simgt{\lower.5ex\hbox{$\; \buildrel > \over \sim \;$}}

\begin{document}
\title{Characterising cosmic birefringence in the presence of galactic foregrounds and instrumental systematic effects}
\author{Baptiste Jost}
 \affiliation{Universit\'{e} de Paris Cit\'e, CNRS, Astroparticule et Cosmologie, F-75013 Paris, France}
 \affiliation{CNRS-UCB International Research Laboratory, Centre Pierre Bin{\'e}truy, IRL2007, CPB-IN2P3, Berkeley, CA 94720, USA}
\author{Josquin Errard}
 \affiliation{Universit\'{e} de Paris Cit\'e, CNRS, Astroparticule et Cosmologie, F-75013 Paris, France}
\author{Radek Stompor}
 \affiliation{CNRS-UCB International Research Laboratory, Centre Pierre Bin{\'e}truy, IRL2007, CPB-IN2P3, Berkeley, CA 94720, USA}
  \affiliation{Universit\'{e} de Paris Cit\'e, CNRS, Astroparticule et Cosmologie, F-75013 Paris, France}

\begin{abstract}
We study a possibility of constraining isotropic cosmic birefringence with help of cosmic microwave background polarisation data in the presence of polarisation angle miscalibration without relying on any assumptions about the Galactic foreground angular power spectra and in particular on their EB correlation.
For this purpose, we propose a new analysis framework based on a generalised parametric component separation approach, which accounts simultaneously on the presence of galactic foregrounds, relevant instrumental effects and external priors.
We find that upcoming multi-frequency CMB data with appropriate calibration priors will allow producing an instrumental-effect-corrected and foreground-cleaned CMB map, which can be used to estimate the isotropic birefringence angle and the tensor-to-scalar ratio, accounting on statistical and systematic uncertainties incurred during the entire procedure.
In particular, in the case of a Simons Observatory-like, three Small Aperture Telescopes, we derive an uncertainty on the birefringence angle of $\sigma(\beta_{b}) = 0.07^\circ$ (0.1$^\circ$), assuming the standard cosmology and calibration priors for all (single) frequency channels with the precision of $\sigma(\alpha_i)= 0.1^\circ$ as aimed at by the near future ground-based multi-frequency experiments. This implies that these experiments could confirm or disprove the recently detected value of $\beta_b=0.35^\circ$ with a significance between $3$ and $5 \sigma$. We furthermore explore the impact of precision of the calibration priors and of foreground complexity on our results and discuss requirements on the calibration precision.
In addition, we also investigate constraints on the tensor-to-scalar ratio, $r$,  which can be derived in the presence of isotropic birefringence and/or polarisation angle miscalibration. We find that the proposed method allows setting constraints on $r$ in such cases, even if no prior is available, and with only a minor increase of the final uncertainty as compared to cases without these effects.
\end{abstract}

\maketitle

\section{Introduction}

Cosmic birefringence rotates the polarisation angle of CMB photons as a consequence of some parity violating mechanism, which could be due to multiple reasons ranging from a violation of the Lorentz symmetry \cite{Carroll1990a} to the impact of specific dark-energy models \cite{Carroll1998} or dark-matter axion-like particles through the Chern-Simons effect \cite{Finelli_2009,Fedderke2019}. This parity violation would lead to non-zero EB correlation, even if the primordial EB correlation is zero as it is the case in the standard cosmology.
Cosmic birefringence can take many forms, depending on the details of the
underlying physical mechanism. It can be isotropic or anisotropic, frequency dependent (similar to Faraday rotation), time dependent or constant. In this paper we focus on the case of time- and frequency- independent, isotropic birefringence.
Hereafter, we will denote the direction-independent birefringence angle as $\beta_b$ and assume no primordial EB correlation, i.e., $C_\ell^{EB,\rm{CMB}}=0$.
The proposed framework is however adaptable to any non-zero primordial EB correlations, as predicted by e.g. anisotropic inflation models \cite{Watanabe_2011} or an asymmetry in primordial GW handedness \cite{Lue_1999, Saito_2007} (chiral gravitational waves).

Detecting cosmic birefringence is of significant scientific importance as it could provide valuable hints about the physics beyond the standard model of particle physics. This, however, requires efficient means of breaking an inherent degeneracy between the birefringence angle and an effective orientation of the polarisation-sensitive CMB detectors with respect to the sky coordinate. Cosmic birefringence and the polarisation angle miscalibration, can also affect the estimation of other cosmological parameters such as the tensor-to-scalar ratio, $r$~\cite{Abitbol_2021}.

Here we propose a self-contained framework, which permits studying both these effects in a consistent and statistically robust manner. We focus on the approach where the degeneracy is broken with the help of calibration constraints assumed to be available for all or some of the single-frequency maps. The way these calibration constraints determine the effective polarisation angle of the recovered CMB map depends on the galactic foregrounds and the details of the instrument design, which we study in the context of the generalised parametric component separation method.

A number of studies have been published recently, addressing some of the issues mentioned above.
The most relevant is the study by \cite{Abitbol_2021} who in the case of the Simons Observatory (SO) estimate the precision with which the polarisation angle needs to be known in order to meet the targeted precision goal on $r$ of $\sigma(r) \approx 10^{-3}$ \cite{Ade_2019}. They obtain the value of a few tenths of a degree. Similarly, for LiteBIRD~\cite{PTEP_LB}, aiming at constraining $r \leq 10^{-3}$, \cite{Vielva_2022} find that the polarisation angle precision should be even more stringent and ranges between an arcminute and a few tens of arcminutes depending on the frequency channel. Our work generalises these approaches by providing a general, self-contained framework for such analyses, allowing to correct for the angle miscalibration to the extent possible, and setting simultaneous constraints on the birefringence and $r$.

From observational perspective, there exists constraints on time-dependent~\cite{PhysRevD.103.042002,PhysRevD.105.022006, SPT-3G:2022ods} and on anisotropic  \cite{Gluscevic_2012,Ade_2017,Ade_2015, Namikawa_2020,Bianchini_2020, Gruppuso_2020} derived, respectively, from BICEP and SPT, and WMAP, POLARBEAR, BICEP Keck, ACT, SPT and Planck, data sets. Such constraints are independent on the knowledge of the absolute polarisation angle as they rely on the variability, temporal or spatial, of the signal.

For isotropic birefringence, \textcite{Minami_method}  proposed to lift the degeneracy between polarisation angle and birefringence by assuming a model for the EB power spectrum of the foregrounds. The method was further elaborated on in \cite{Diego_Palazuelos_2022}.
With this model it is possible to estimate the polarisation angle of the telescope by fitting the parameters of the foreground EB to observations.
The first results using this method applied to the Planck data sets are very promising and potentially hint at non-zero isotropic birefringence with $\beta_b = 0.35^\circ \pm 0.14^\circ$ \cite{Minami_2020}.  \textcite{Diego_Palazuelos_2022} found similar results but had to assume more complex foreground models such as filament models from \cite{Clark_2021} and \cite{Huffenberger_2020} as well as the COMMANDER sky model \cite{Planck2020} to ensure that the found value of birefringence angle is independent on the area of observed sky. Even more recently \textcite{Bire_Planck_WMAP} used this method on the Planck and WMAP data and found $\beta_b = 0.37^\circ \pm 0.14^\circ$ with $f_{sky}=0.62$ and
$\beta_b = 0.342^\circ \vphantom{.}^{+0.094^\circ}_{-0.091^\circ}$ with $f_{sky}=0.92$. They also found the constraints consistent with a frequency independent birefringence angle in agreement with the Planck PR4 data only analysis~ \cite{Eskilt_2022}.  While these results potentially hint at the non-zero birefringence, they all rely on assumptions about the EB angular power spectrum of the foregrounds for which reliable models or measurements are lacking at this time.
The approach discussed here is thus complementary to these studies.

Other relevant methods include the so-called self-calibration method of~\textcite{Keating_2012} and the birefringence tomography proposed in~\textcite{Sherwin2021}.
The self-calibration method constrains the polarisation angle so that the EB correlations of the signal contained in the map vanishes.
This allows in principle to calibrate the angle with high precision but, by assumption, it rules out any possible detection of isotropic cosmic birefringence.
The birefringence tomography aims at measuring the difference between a birefringence angle induced at recombination and one at reionisation (corresponding respectively to small and large angular scales of the CMB spectra) therefore constraining the change of the birefringence angle between the two corresponding redshifts.
Both these methods should ideally be applied to the foreground-cleaned CMB maps as produced at the outcome of the component separation procedures. For multifrequency observations with potentially different miscalibration angles at different frequencies, this however requires understanding the impact of such effects on the component separation and the expected level of foreground residuals in the recovered maps. This work is therefore also relevant for these methods.

In this paper, we first generalise a parametric component separation method as applied to a multi-frequency set of Stokes maps so it can correct jointly for instrumental effects and cleaning complex foregrounds. We use calibrations of the polarisation angle of the telescopes to break the polarisation angle-birefringence degeneracy, whatever are the spatial properties of the foregrounds or whether cosmic birefringence is present or not. We then assess the impact of this procedure on the joint estimation of the birefringence angle, $\beta_b$, and the tensor-to-scalar ratio, $r$. This approach can be seen as a generalisation of the self-calibration method~\cite{Keating_2012} extended to allow for a detection of isotropic birefringence, simultaneously with the amplitude of the primordial gravitational waves, and explicitly accounting on the foreground contaminations.

While the presented approach should be eventually implemented within an actual CMB data analysis pipeline, hereafter we recast it as a forecasting tool in order to derive realistic and robust, ensemble-averaged constraints on the cosmological parameters, $\beta_b$ and $r$, and to provide meaningful precision requirements for the calibration priors.

\section{Method}

Our method is composed of two steps. The first step consists in a simultaneous estimation of foreground and instrumental parameters performed as part of the generalised parametric component separation described in~\cite{stompor2016}. It uses a generalised version of the so-called spectral likelihood and yields constraints on foreground and instrumental parameters. These are then used to derive estimates of the sky components including that of the CMB, as well as their generalised statistical uncertainties.

The second step then constraints cosmological parameters from the CMB maps and their covariances as derived on the first step. This is done with help of the likelihood obtained assuming that CMB signal is (nearly) Gaussian and isotropic and its covariance is given by the CMB power spectra, which in turn depend on cosmological parameters. Consequently, the covariance model does not account for the presence of foreground residuals in the CMB map estimated on the first step.  This may then lead to systematic errors in the estimated parameters, which we study hereafter, together with their statistical uncertainties.

We describe the entire procedure in detail below.

\subsection{Data Model}
The input for the component separation step are single frequency maps reconstructed from actual measurements of a CMB experiment.
These maps are gathered in a single data vector $\boldsymbol{d}$; for each sky pixel we store the measured sky amplitudes for each frequency, so $\boldsymbol{d}$ contains $n_\mathrm{stokes}\times n_f$ maps.
We model $\boldsymbol{d}$ as,
\begin{equation} \label{eq:multi_pix_data}
\boldsymbol{d} = \boldsymbol{X} \, \boldsymbol{A} \, \boldsymbol{\mathcal{B}} \, \boldsymbol{c} + \boldsymbol{n}
\end{equation}
where $\boldsymbol{n}$ stands for the noise in all maps concatenated together. $\boldsymbol{c}$ is a vector composed of all maps of the sky signals. It contains therefore $n_\mathrm{stokes}\times n_{\rm{comp}}$ maps. $\boldsymbol{\mathcal{B}}$ acts on $\boldsymbol{c}$ and models the impact of birefringence on the CMB.
$\boldsymbol{A}$ is the mixing matrix that acts on the component vector synthesising them into the corresponding frequency maps for each observation channel of the considered telescope.

Finally $\boldsymbol{X}$ is the instrumental response matrix that models how the instrument's characteristics may affect the sky signal at each frequency.

We define the mixing matrix for each sky pixel $p$ as $\boldsymbol{A}_p(\{\beta_{fg}\})$. It scales the sky component amplitudes in frequency and co-adds them together. Each column of $\boldsymbol{A}$ corresponds to a component and each row to an observation frequency for each considered Stokes parameter, either $Q$ or $U$ in the following. The scaling of foreground components with respect to frequency is assumed to be parameterised by a set of spectral parameters $\{\beta_{fg}\}$. We adopt the CMB units so that the elements of $\boldsymbol{A}$ acting on the CMB sky signal are all equal to $1$. The mixing matrix has $(n_\mathrm{stokes}\times n_f) \times (n_\mathrm{stokes}\times n_{comp})$ elements. Here the mixing matrix does not mix between different Stokes parameters and corresponding elements of the mixing matrix vanish, leaving only $n_\mathrm{stokes}\times n_f\times n_{comp}$ non-zero elements, and the elements of $\boldsymbol{A}$ corresponding to $Q$ \& $U$ are equal.
\\
Given that our focus is on isotropic cosmic birefringence $\boldsymbol{\mathcal{B}}$, we consider polarisation angle misalignment as the only instrumental effects. Other effects should be considered in a more complete analysis as discussed in Sect.~\ref{sec:conclusions}. We consider only $Q$ and $U$ Stokes parameter maps, both on the input and the output of the component separation procedure. Therefore, from now on $n_\mathrm{stokes}=2$. The formalism, which follows, can be straightforwardly generalised to include any combination of the Stokes parameters, if needed. The general rotation matrix in $Q$ \& $U$ space for the spin 2 polarisation vector is denoted $\boldsymbol{\mathcal{R}}(\alpha)$ and is defined as\footnote{Note that some references use the other sign convention for the rotation matrix, however this does not affect the results presented in this work.},
\begin{equation}
\label{eq:def_rotation}
\begin{split}
     \begin{pmatrix}
    Q_{out} \\
    U_{out} \\
\end{pmatrix}
&=
\begin{pmatrix}
    \cos(2\alpha) &    \sin(2\alpha) \\
    -\sin(2\alpha) & \cos(2\alpha)
\end{pmatrix}
 \begin{pmatrix}
    Q_{in} \\
    U_{in} \\
\end{pmatrix} \\
&\equiv \boldsymbol{\mathcal{R}}(\alpha) \begin{pmatrix}
    Q_{in} \\
    U_{in} \\
\end{pmatrix}.
\end{split}
\end{equation}

Effects of cosmic birefringence at the map level can be represented as a block diagonal matrix, with each block corresponding to a different sky pixel. As birefringence acts only on the CMB $Q$ \& $U$ elements the CMB-CMB subblock of each block of $\boldsymbol{\mathcal{B}}$ is a rotation matrix $\boldsymbol{\mathcal{R}}(\beta_b)$, where $\beta_b$ the birefringence angle. The other components are unchanged by $\boldsymbol{\mathcal{B}}$ and the remaining subblocks of the matrix are equal to the identity matrix.
For definiteness, we will assume that properties of the ``primordial'' CMB, i.e., as contained in the component vector, $\mathbf{c}$, are well-defined and known, and for simplicity throughout this paper, we will take the ``primordial'' CMB EB cross-correlation to be zero. This makes the definition of the birefringence angle (in the absence of other effects studied below) well defined. However, incorporating the cases with some specific, non-vanishing, ``primordial'' EB correlations is straightforward. We redefine the sky signal to an effective sky signal after birefringence as: $\boldsymbol{s}_p \equiv \boldsymbol{\mathcal{B}}(\beta_b) \, \boldsymbol{c}_p$.
We also note that the framework could be generalised to the anisotropic case by simply allowing the birefringence angle to vary between sky pixels.\\
For simplicity we model polarisation miscalibration with a single angle for each single frequency map. This can be modified as needed depending on specific experiment conditions, assigning one angle per focal plane wafer or on the contrary using one angle per multi-frequency instrument, for instance. The miscalibration of polarisation angles is described as a rotation matrix acting on each pixel of $Q$ \& $U$ maps of a particular frequency channel.
The corresponding instrumental response matrix, $\boldsymbol{X}(\{\alpha_1,...,\alpha_{n_f}\})$, is then a block diagonal matrix with a block assigned to each sky pixel and composed of frequency-specific sublocks given by a rotation matrix $\boldsymbol{\mathcal{R}}(\alpha_i)$ acting on the $Q$ \& $U$ Stokes parameters of the $i^{th}$ frequency channel of the vector given by $\boldsymbol{A}_p(\{\beta_{fg}\}) \, \boldsymbol{\mathcal{B}}(\beta_b) \, c_p$.
\\
The miscalibration angles are handled at the same time as foreground spectral indices.
We therefore introduce an effective mixing matrix defined as $\boldsymbol{\Lambda}_p(\{\Gamma \} ) \equiv \boldsymbol{X}(\{\alpha_1,...,\alpha_{n_f}\}) \, \boldsymbol{A}_p(\{\beta_{fg}\})$, where $\{\Gamma \} \equiv \{ \{\alpha\},\{\beta_{fg}\}\}$ and $\{\alpha\}$ denotes $\{\alpha_1,...,\alpha_{n_f}\} $. A generic element of $\{\Gamma \}$ will be referred to as $\gamma$.
We can now rewrite the data model in Eq.~\ref{eq:data_model} as,
\begin{equation}
\begin{split}
    \boldsymbol{d}_p &=\underbrace{ \boldsymbol{X}_p(\{\alpha_1,...,\alpha_{n_f}\}) \,
    \boldsymbol{A}_p(\{\beta_{fg}\})}_{\boldsymbol{\Lambda}_p(\{\Gamma \})} \,
    \underbrace{\boldsymbol{\mathcal{B}}(\beta_b)\, \boldsymbol{c}_p}_{\boldsymbol{s}_p} + \boldsymbol{n}_p\\
\end{split}
\label{eq:data_model}
\end{equation}
In this perspective the instrumental and foreground parameters are both merely unknowns of a global fitting problem, however, the manner in which they impact the entire procedure is rather different. The foreground parameters are sky-component specific and depend on the assumed model, hence they do not depend on the number of available frequency channels. More channels permits in general better estimation of the foreground parameters. This is not always so for the instrumental parameters, as they are commonly specific to frequency channels and more frequencies typically mean more parameters. This is for instance the case of the polarisation angle misalignment as discussed in detail later in this work. (See \cite{verges2020} for a counter example). In such cases increasing the number of available channels may not improve the problem's stability and instead other means, such as priors, may need to be incorporated in the component separation formalism.

Let us consider an arbitrary instrumental angle $\alpha_0$, we can always write,
\begin{eqnarray}
\boldsymbol{X}_p(\{\alpha_1,...,\alpha_{n_f}\}) & = & \mathbf{X}(\{\alpha_1-\alpha_0, ..., \alpha_{n_f}-\alpha_0\})\nonumber\\
&\times& \boldsymbol{X}_p(\{\alpha_0, ..., \alpha_{0}\} \\
& = & \boldsymbol{X}_p(\{\alpha'_1,...,\alpha'_{n_f}\})\,\boldsymbol{X}_p(\{\alpha_0, ..., \alpha_{0}\}),\nonumber
\end{eqnarray}
where all the angles with a prime include an extra common rotation by an angle $-\alpha_0$ and the rightmost factor in the last equation rotates all channels by the same angle $\alpha_0$ to compensate for this. We note that rotating all frequency channels by the same angle is equivalent to rotating all sky components by the very same angle, i.e.,
\begin{eqnarray}
\boldsymbol{X}_p(\{\alpha_0, ..., \alpha_{0}\})\,
\boldsymbol{A}_p\,\mathbf{s}_p & = &
\boldsymbol{A}_p\,
\boldsymbol{\tilde X}_p(\{\alpha_0, ..., \alpha_{0}\})\, \mathbf{s}_p,
\end{eqnarray}
where $\boldsymbol{\tilde X}_p$ is a rotation operator analogous to $\boldsymbol{X}_p$ but operating on the sky components instead of the frequency channels and for definiteness we assume, as always, that CMB is the first component. We can therefore rewrite the signal term of our data model in Eq.~\ref{eq:data_model} as,
\begin{align}
\boldsymbol{A}_p\,
\boldsymbol{\tilde X}_p(\{\alpha_0, ..., \alpha_{0}\})\,
\boldsymbol{\mathcal{B}}(\beta_b) \, \mathbf{c}_p & = \nonumber \\
= \;  \boldsymbol{A}_p\,
\boldsymbol{\tilde X}_p(\{&\alpha_0+\beta_b, ..., \alpha_{0}\})\, \mathbf{c}_p
\label{eqn:deg}
\\
= \;
\boldsymbol{A}_p\,
\boldsymbol{\mathcal{B}}(\alpha_0&+\beta_b)\, \mathbf{c'}_p.
\nonumber
\end{align}
The new sky components, $\mathbf{c'}$, defined above, contain the same CMB signal as $\mathbf{c}$, and, in particular, its EB correlation of the CMB signal continues to vanish as we require throughout here. The foregrounds signals are however modified due to the rotation by the angle $\alpha_0$.
In the absence of any additional assumptions both $\mathbf{c}$ and $\mathbf{c'}$ and the corresponding birefringence angles, $\beta_b$ and $\beta_b+\alpha_0$, provide a legitimate solution to the problem consistent with the data, $\mathbf{d}$. Consequently, the problem does not have a unique solution for foreground components and the birefringence angle.
In practice this implies that employing any of the standard methods of solving the inverse problem in Eq.~\ref{eq:data_model} is going to have a degeneracy and that we will need  some additional assumptions to break it. These extra assumptions could concern any of the foreground components, e.g., by defining their EB cross-correlations, as done, for instance, in the approaches of~\cite{Minami_method, Diego_Palazuelos_2022}, or provide some external constraints on the common rotation angle, $\alpha_0$. While both of these could be considered in the framework proposed here, this is the second option we focus on in the following.

\subsection{Joint Parametric Component Separation and Systematic Effects Correction}
\label{sub:comp_sep_method}

\subsubsection{Parameter Estimation}
We adapt the fiducial parametric component separation methods of \textcite{stompor2016} to take into account the generalised data model of Eq.~\ref{eq:data_model}.
Replacing the standard mixing matrix of \cite{stompor2016} with the effective mixing matrix $\boldsymbol{\Lambda}_p$ and the sky signal vector with the effective one $s_p$ allows us to jointly fit for systematic effects and foreground parameters.
Depending on the considered systematic effects some parameters might be degenerate, such as the absolute polarisation angles of the detectors. As mentioned before, some of these degeneracies can be lifted using calibration priors. But the impact of those on the statistical error of the parameter estimation needs to be assessed and propagated correctly throughout the pipeline.
\\
As in \textcite{verges2020}, adapting the spectral likelihood maximised over sky signals from \cite{stompor2016} results in the following log-likelihood,
\begin{equation}
\begin{split}
        S &\equiv -2 \ln(\mathcal{L}(\{\Gamma \})  = \text{cst} \, + \\
        & - \sum_{p}{\text{tr} \left( \boldsymbol{N}_p^{-1}\boldsymbol{\Lambda}_p (\boldsymbol{\Lambda}_p^t \boldsymbol{N}_p^{-1} \boldsymbol{\Lambda}_p )^{-1}\boldsymbol{\Lambda}_p^t \boldsymbol{N}_p^{-1} \boldsymbol{d}_p \boldsymbol{d}_p^t \right)}    \label{eq:spectral_like}
\end{split}
\end{equation}
where $\boldsymbol{N}_p$ is the noise covariance matrix. Optimising this likelihood gives us an estimation of both foreground parameters $\beta_{fg}$ and instrumental parameters $\{\alpha\}$.
This is the likelihood we would use while analysing a specific, actual or simulated, data set.
In the forecasting procedure we average the likelihood over both CMB and noise realisation similarly as in \cite{stompor2016},
\begin{equation}
    \label{eq:forecasting_like}
    \langle S \rangle = -\sum_{p}{ \text{tr} \left( (\boldsymbol{N}_p^{-1} - \boldsymbol{P}_p( \{\Gamma \})) \langle \boldsymbol{d}_p \boldsymbol{d}_p^t \rangle \right)}
\end{equation}
where $\boldsymbol{P}_p$ is the projection operator defined as,
\begin{equation}
    \label{eq:projection}
    \boldsymbol{P}_p(\{ \Gamma \}) \equiv \boldsymbol{N}_p^{-1} -  \boldsymbol{N}_p^{-1}\boldsymbol{\Lambda}_p (\boldsymbol{\Lambda}_p^t \boldsymbol{N}_p^{-1} \boldsymbol{\Lambda}_p )^{-1}\boldsymbol{\Lambda}_p^t \boldsymbol{N}_p^{-1},
\end{equation}
and $\langle \boldsymbol{d}_p \boldsymbol{d}_p^t \rangle$ is given by:
\begin{equation}
\label{eq:ddt}
\begin{split}
    \langle \boldsymbol{d}_p \boldsymbol{d}_p^t \rangle &= \langle \boldsymbol{\hat{\Lambda}}_p \boldsymbol{\hat{s}}_p \boldsymbol{\hat{s}}_p^t \boldsymbol{\hat{\Lambda}}_p^t \rangle + \langle \boldsymbol{n}_p \boldsymbol{n}_p^t \rangle \\
    &= \boldsymbol{\Lambda}_p(\{ \hat{\Gamma} \}) \langle \boldsymbol{\hat{s}}_p \boldsymbol{\hat{s}}_p^t \rangle \boldsymbol{\Lambda}^t_p( \{ \hat{\Gamma} \} ) + \boldsymbol{N}_p.
\end{split}
\end{equation}
In Eq.~\ref{eq:ddt}, $\boldsymbol{\hat{\Lambda}}_p$ is $\boldsymbol{\Lambda}_p$ evaluated at the true values of parameters $\{ \hat{\Gamma} \}$ for both instrumental parameters $\{\hat{\alpha}\}$ and foreground parameters $\{\hat{\beta}_{fg}\}$ and
$\boldsymbol{\hat{s}}_p$ is the true effective sky signal containing CMB and foregrounds. \\
We can express $\boldsymbol{\hat{\Lambda}}_p$ and $\boldsymbol{\hat{s}}_p$ in such a way as to distinguish between the CMB terms and the foregrounds term in Eq.~\ref{eq:ddt}. $\boldsymbol{\hat{s}}_p^{\rm{cmb}}$ represents the two rows of $\boldsymbol{\hat{s}}_p$ that correspond to the $Q$ \& $U$ CMB signals, and $\boldsymbol{\hat{s}}_p^{\rm{fg}}$ to the leftover components. Similarly $\boldsymbol{\hat{\Lambda}}^{\rm{cmb}}_p$ is made of the two columns of $\boldsymbol{\hat{\Lambda}}_p$ that act on the CMB components in $\boldsymbol{\hat{s}}_p$, and $\boldsymbol{\hat{\Lambda}}^{\rm{fg}}_p$ the other columns. Eq.~\ref{eq:ddt} becomes,
\begin{equation}
\begin{split}
\langle \boldsymbol{d}_p \boldsymbol{d}_p^{\rm{t}} \rangle &=  \boldsymbol{\hat{\Lambda}}_p^{\rm{cmb}} \langle \boldsymbol{\hat{s}}^{\rm{cmb}}_p \boldsymbol{\hat{s}}_p^{\rm{cmb,t}} \rangle  \boldsymbol{\hat{\Lambda}}_p^{\rm{cmb,t}} \\
&+ \boldsymbol{\hat{\Lambda}}^{\rm{fg}}_p  \boldsymbol{\hat{s}}^{\rm{fg}}_p \boldsymbol{\hat{s}}_p^{\rm{fg,t}}  \boldsymbol{\hat{\Lambda}}^{\rm{fg,t}}_p + \boldsymbol{N}_p.
\end{split}
\end{equation}
This expression can be straightforwardly generalised to the cases of non-parameterisable foreground models by simply replacing $\boldsymbol{\hat{\Lambda}}^{\rm{fg}}_p  \boldsymbol{\hat{s}}^{\rm{fg}}_p$ by a vector of foreground signals at each frequency modified by the instrumental effects operator computed for the true values of the instrumental parameters, i.e., $\mathbf{\hat{X}}_p\,\boldsymbol{\hat{f}}_p$. The \emph{parametric} model used in the effective mixing matrix $\boldsymbol{\Lambda}_p$ would then mismatch with the data and could lead to foreground residuals and a bias in the cosmological parameters.
The average over CMB and noise realisations does not affect $\boldsymbol{\hat{f}}_p$ and we use the output frequency maps from \texttt{PySM} to compute the $\boldsymbol{\hat{f}}_p \boldsymbol{\hat{f}}_p^{\rm{t}}$ term.

\begin{figure}
    \centering
    \includegraphics[width=\linewidth]{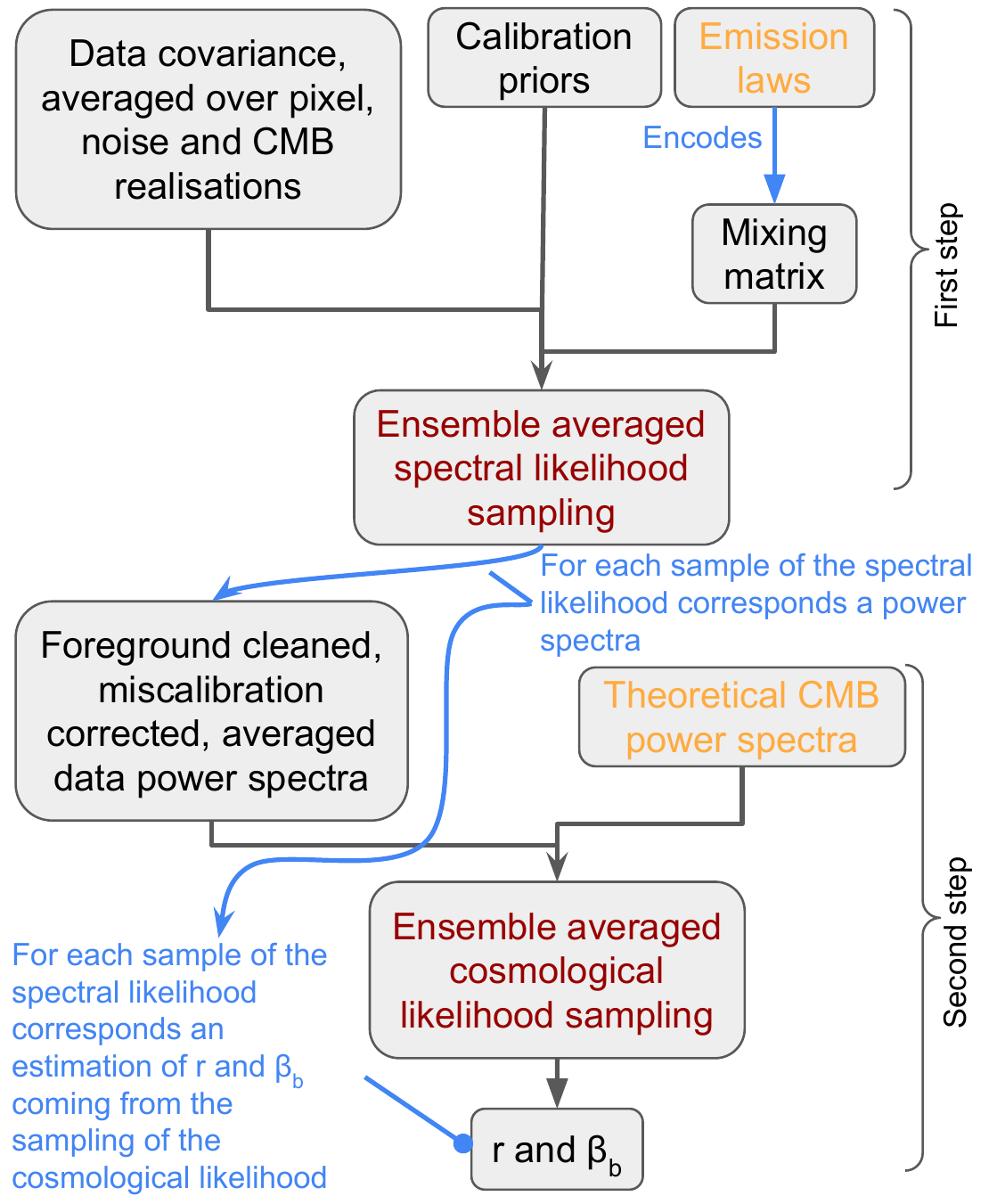}
    \caption{Diagram depicting the main features of the method in its forecasting rendition. In red there are the two main steps of the algorithm, in orange -- the basic assumptions going in constructing the likelihoods, and in black -- the various inputs and outputs of each of the main steps.}
    \label{fig:method_diagram}
\end{figure}

The CMB term can be expressed as an average over pixel of the correlation between the $Q$ and $U$ CMB signals. Here we focus on the case where the effective mixing matrix $\boldsymbol{\Lambda}_p$, the instrument matrix, $\boldsymbol{\hat{X}}_p$, and the noise covariance matrix $\boldsymbol{N}_p$ are all pixel independent. We can then rewrite Eq.~\ref{eq:forecasting_like} as,
\begin{widetext}
\begin{eqnarray}
\centering
    \langle S \rangle &=& - \text{tr} \left\{ (\boldsymbol{N}^{-1} - \boldsymbol{P})
    \left(n_{\rm{pix}}\boldsymbol{N} + \boldsymbol{\hat{\Lambda}}^{\rm{cmb}} \left( \sum_{p}{\langle \boldsymbol{\hat{s}}^{\rm{cmb}}_p  \boldsymbol{\hat{s}}_p^{\rm{cmb,t}} \rangle} \right) \boldsymbol{\hat{\Lambda}}^{\rm{cmb,t}} +
    \boldsymbol{\hat{X}} \sum_{p}{\boldsymbol{\hat{f}}_p \boldsymbol{\hat{f}}_p^{\rm{t}}} \boldsymbol{\hat{X}}^{\rm{t}} \right) \right\}
     \\
\label{eq:average_spectral}
    &=& - \text{tr} \left\{n_{\rm{pix}} (\boldsymbol{N}^{-1} - \boldsymbol{P})
    \left(\boldsymbol{N} + \boldsymbol{\hat{\Lambda}}^{\rm{cmb}} \boldsymbol{S}^{\rm{cmb}} \boldsymbol{\hat{\Lambda}}^{\rm{CMB,t}} + \boldsymbol{\hat{X}} \boldsymbol{F} \boldsymbol{\hat{X}}^{\rm{t}} \right)\right\}
\end{eqnarray}
\end{widetext}
where $n_{\rm{pix}}$ is the total number of observed pixels over which the summation acts.
We denote the pixel averages of the CMB sky component as $\boldsymbol{S}^{\rm{cmb}}$ and of foregrounds frequency maps as $\boldsymbol{F}$:
\begin{eqnarray}
\centering
\label{eq:S_CMB}
    \boldsymbol{S}^{\rm{cmb}} &\equiv& \frac{1}{n_{\rm{pix}}}\sum_{p}{\langle \boldsymbol{\hat{s}}^{\rm{cmb}}_p \boldsymbol{\hat{s}}_p^{\rm{cmb,t}}\rangle}\\
\label{eq:F_fg}
    \boldsymbol{F} &\equiv& \frac{1}{n_{\rm{pix}}}\sum_{p}{\boldsymbol{\hat{f}}_p \boldsymbol{\hat{f}}_p^{\rm{t}}}
\end{eqnarray}
Only the projection matrix $\boldsymbol{P}$ needs to be updated when exploring the likelihood, Eq.~\ref{eq:average_spectral},  which makes it more efficient to explore the parameter space $\{\Gamma \}$.
We refer to the objects defined in Eqs.~\ref{eq:S_CMB} and~\ref{eq:F_fg}, as signal
covariances and together with the noise covariance, they provide a complete and necessary description of the input data which is needed by our forecasting pipeline as shown in Fig.~\ref{fig:method_diagram}.

Depending on the number and specific nature of instrumental and foreground parameters considered in the problem, some degeneracies between the parameters may arise, see e.g., Eq.~\ref{eqn:deg} and a discussion there, preventing a robust determination of at least some of them. To deal with those, prior knowledge of some parameters may be required. It can come from instrumental calibration for instance.
We allow for priors on the miscalibration angles, $\{\alpha_i\}$, and assume them to be Gaussians with a mean, $\Tilde{\alpha}_i$, corresponding to the value of the parameter as measured during the calibration campaign and the dispersion, $\sigma_{\alpha_i}$, reflecting the error of the measurement. This can be straightforwardly generalised to other parameters.
The complete log-likelihood is then given by:
\begin{equation}
\label{eq:spec_like_prior_notAver}
    S' \equiv \langle S \rangle + \sum_{\alpha_i}{\frac{(\alpha_i - \Tilde{\alpha}_i)^2}{\sigma_{\alpha_i}^2}}.
\end{equation}

In the following we will assume that our priors are unbiased in a sense that if multiple calibration campaigns were to be performed the best-fit values obtained from each measurement will be drawn from a Gaussian with the mean corresponding to the true value of the parameter and the dispersion set by the measurement error. We will then average our likelihood over the ensemble of the calibration procedures. The effective log-likelihood we will use hereafter therefore reads,
\begin{equation}
\label{eq:spec_like_prior}
    S' \equiv \langle S \rangle + \sum_{\alpha_i}{\frac{(\alpha_i - \hat{\alpha}_i)^2}{2\sigma_{\alpha_i}^2}}.
\end{equation}

\subsubsection{Residuals}
For each set of sampled spectral and instrumental parameters $\Gamma$ we compute the effective mixing matrix $\boldsymbol{\Lambda}$ and use it to get the estimate of the separated sky components,
\begin{equation}
\label{eq:recMap}
\begin{split}
   \boldsymbol{m}_p &= (\boldsymbol{\Lambda}^t \boldsymbol{N}^{-1} \boldsymbol{\Lambda} )^{-1}\boldsymbol{\Lambda}^t \boldsymbol{N}^{-1} \, \boldsymbol{d}_p
   \equiv \boldsymbol{W}_p \, \boldsymbol{d}_p \\
   & = \boldsymbol{W}_p \, \boldsymbol{\hat{\Lambda}}^\mathrm{cmb} \, \boldsymbol{\hat{s}}^\mathrm{cmb}_p\,+\,\boldsymbol{W}_p\boldsymbol{\hat{X}}\,\boldsymbol{\hat{f}}_p\,+\,\boldsymbol{n}_p\\
   & = \boldsymbol{W}_p \, \boldsymbol{\hat{\Lambda}} \, \boldsymbol{\hat{s}}_p\,+\,\boldsymbol{n}_p,
\end{split}
\end{equation}
where the last equality is only true if the foreground signal at the required frequencies can be modelled as a linear combination of the foreground component templates.
The sought-after CMB map corresponds then to the first element of the sky component estimates, $\boldsymbol{m}$. The latter provides an unbiased (over the statistical ensemble of noise realisations) estimate of the true sky components if $\boldsymbol{W}_p \, \boldsymbol{\hat{\Lambda}} = 1$ for all pixels $p$. This will be in general only true if the foregrounds can indeed be modelled as a linear combination of some templates (as in the last line of Eq.~\ref{eq:recMap}), and $\boldsymbol{\Lambda} = \boldsymbol{\hat{\Lambda}}$.
However, if $\boldsymbol{\Lambda}$ needs to be recovered from data then, in the best case, the equality above will hold only on average and, case-by-case, the estimates of the sky components will include contributions from the others.
Hereafter we refer to these additional contributions as residuals.

We split the component vector, $\boldsymbol{s}$, the mixing matrix, $\boldsymbol{\Lambda}$, and the map-making operator, $\boldsymbol{W}_p$ into a CMB and foreground parts. For $\boldsymbol{W}_p$ the split is performed row-wise.
We can then express noise-free CMB map estimate as:
\begin{equation}
\label{eq:scmb_estimation}
    \boldsymbol{s}^{\rm{cmb}}_p
    = \boldsymbol{W}^{\rm{cmb}}\left(\boldsymbol{\hat{\Lambda}}^{\rm{cmb}} \boldsymbol{\hat{s}}_p^{\rm{cmb}}
    + \boldsymbol{\hat{X}}\, \boldsymbol{\hat{f}}_p\right)
\end{equation}
Note again that here $\boldsymbol{s}^{\rm{cmb}}_p$ has two elements corresponding to two Stokes parameters.

As highlighted by Eq.~\ref{eq:scmb_estimation},
there are two types of effects which affect the estimation of the CMB map:
\begin{enumerate}
   \item[-] A \textbf{multiplicative effect} coming from the action of $\boldsymbol{W}^{\rm{cmb}} \boldsymbol{\hat{\Lambda}}^{\rm{cmb}}$ on $\boldsymbol{\hat{s}}_p^{\rm{cmb}}$. Without miscalibration $\boldsymbol{W}^{\rm{cmb}} \boldsymbol{\hat{\Lambda}}^{\rm{cmb}}$ would be the identity. However, this is not so here due to the action of the instrumental response matrix $\boldsymbol{X}$. Consequently, and unlike in~\cite{stompor2016}, in our case not all of the actual CMB signal is bound to end up in the CMB map estimate. Instead, part of it may leak to the estimates of the other components and the CMB signal found in the CMB estimate may be corrupted.
    \item[-] An \textbf{additive effect} coming from the contamination of foregrounds in $\boldsymbol{W}^{\rm{cmb}} \boldsymbol{\hat{X}} \boldsymbol{\hat{f}}_p$ which is closely related to the usual definition of residuals in the context of parametric component separation.
\end{enumerate}
Both these effects will in general give rise to a residual in the recovered CMB map either by adding some spurious foreground signal -- the additive effect -- or by directly misestimating the CMB signal -- the multiplicative effect. This residual can subsequently potentially bias the estimation of cosmological parameters. We note that due to the presence of the multiplicative effect, related to the inclusion of the instrumental effects, the expression for the residual becomes more complex in our case than it was in the original formalism of~\cite{stompor2016}, see, e.g.,~\cite{verges2020}. This can potentially make a direct generalisation of that formalism more cumbersome.
In the formalism proposed hereafter we therefore do not perform Taylor expansion of the residuals with respect to the parameters. Instead, while computing the data matrix, we compute analytically only the term due to the multiplicative effect while the additive effect is computed numerically as in Eq.~\ref{eq:scmb_estimation} for each sample of instrumental and foreground parameters and averaging over these is performed with the help of sampling of the spectral likelihood. This simplifies the formalism significantly and makes no assumption that the errors on spectral parameters derived from the spectral likelihood are small.
 A possible downside of this approach is that we lose some insight into the morphology and sources of the residuals.

\subsection{Cosmological Parameter Estimation}

\begin{figure*}
\centering
\makebox[\textwidth][c]{\includegraphics[width=\paperwidth]{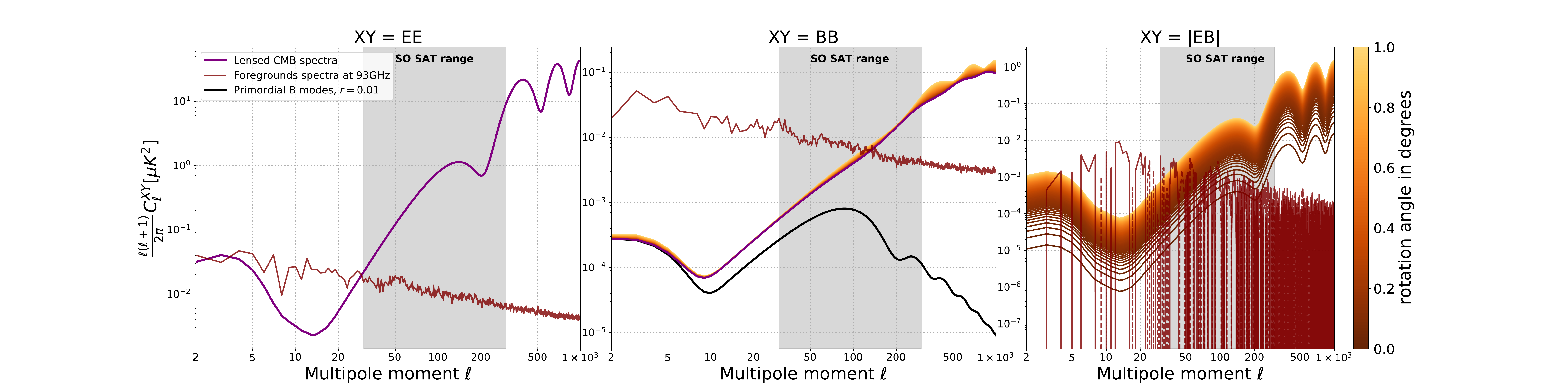}}
\caption{Impact of an isotropic birefringence angle on the CMB lensed spectra. The galactic foregrounds power spectra, unaffected by birefringence and estimated here at $93$GHz on a $f_{\rm sky}\approx 10\%$ SO SAT-like patch~\cite{Ade_2019}, are represented in dark-red.}
\label{fig:spectra}
\end{figure*}

We use the cleaned and corrected CMB map obtained in the previous step, Eq.~\ref{eq:recMap}, in order to estimate the cosmological parameters that we are interested in using the standard cosmological likelihood.
The estimated CMB maps consist of multiple contributions as detailed in Eq.~\ref{eq:scmb_estimation}. We use it then to characterise the statistical properties of the CMB maps averaged over the ensemble of CMB and noise realisations, thus including the effects due to the component separation. These are then used to construct the ensemble averaged cosmological likelihood, which is subsequently used to derive forecasts concerning cosmological parameters.

In the reminder of this section  we detail the procedure and mathematical framework in its most general rendition, specialising it to the case of the joint estimation of the tensor-to-scalar ratio, $r$, and the cosmic birefringence angle, $\beta_b$, only at the end.

\subsubsection{CMB Covariances in Harmonic Domain}
We build the cosmological likelihood in the harmonic domain as this is convenient for the purpose of the forecasting pipeline. However, the analogous constructions can be performed in the pixel domain as could be more appropriate for actual, case-by-case applications of the procedure. In the harmonic domain, we represent sky maps via their harmonic expansion coefficients denoted as $\boldsymbol{a}_j$, where $j$ is related to the multipole numbers $(\ell,m)$ as $j \equiv \ell^2 + \ell + m$. As before, we use a hat to distinguish the true values from the estimates. We collect the harmonic coefficients for the combined foreground signals for all frequency bands in a single vector denoted, $\boldsymbol{a}^\mathrm{fg}$, and those of the CMB signals into a single vector $\boldsymbol{a}^\mathrm{cmb}$. The CMB signal estimate in the harmonic domain after the generalised component separation is then computed \emph{case by case} for each set of values of $\Gamma$, e.g., a sample from the generalised spectral likelihood as in the formalism described here, is given by:
\begin{equation}
\begin{split}
    \label{eq:harmonic_map}
    \boldsymbol{a}_j^{\rm{cmb}} &= \boldsymbol{W}^{\rm{cmb}} \boldsymbol{\hat{\Lambda}}^{\rm{cmb}} \boldsymbol{\hat{a}}^{\rm{cmb}}_j + \boldsymbol{W}^{\rm{cmb}} \boldsymbol{\hat{X}} \boldsymbol{\hat{a}}^\mathrm{fg}_j \\
    &+ \boldsymbol{W}^{\rm{cmb}} \boldsymbol{\hat{a}}^{\rm{noise}}_j
\end{split}
\end{equation}
Our cosmological likelihood is averaged over the CMB and noise realisations and therefore uses the covariance of the recovered CMB map data after generalised component separation defined as $\boldsymbol{E}_{jj'} \equiv \langle \boldsymbol{a}_{j} \boldsymbol{a}_{j'}^{t} \rangle$. Using Eq. \ref{eq:harmonic_map} and assuming that there is no cross-correlations between $\boldsymbol{a}^{\rm{cmb}}_j$, $\boldsymbol{a}^{\rm{fg}}_{j}$ and $\boldsymbol{a}^{\rm{noise}}_{j}$, we get:
\begin{equation}
    \label{eq:Ejj}
    \begin{split}
    \boldsymbol{E}_{jj'} &=     \boldsymbol{{W}}^{\rm{cmb}} \boldsymbol{\hat{\Lambda}}^{\rm{cmb}} \boldsymbol{\mathcal{C}}^{\rm{cmb}}_\ell \delta_{jj'}\boldsymbol{\hat{\Lambda}}^{\rm{cmb,t}} \boldsymbol{{W}}^{\rm{cmb,t}} \\
    &+
    \boldsymbol{{W}}^{\rm{cmb}} \boldsymbol{\hat{X}}
    \boldsymbol{\hat{a}}^{\rm{fg}}_j \boldsymbol{\hat{a}}^{\rm{fg,t}}_{j'}
    \boldsymbol{\hat{X}}^{\rm{t}} \boldsymbol{{W}}^{\rm{cmb,t}} \\
    &+ \boldsymbol{\mathcal{C}}^{\rm{noise}}_\ell\delta_{jj'}
\end{split}
\end{equation}
where $\boldsymbol{\mathcal{C}}^{\rm{cmb}}_\ell  \equiv \frac{1}{2\ell+1}\sum_{m} \langle \boldsymbol{\hat{a}}^{\rm{cmb}}_j \boldsymbol{\hat{a}}^{\rm{cmb,t}}_{j}\rangle$,
and $\boldsymbol{\mathcal{C}}^{\rm{noise}}_\ell$ stand for the CMB and noise spectra respectively. And as we consider $Q$ \& $U$ polarisation information only, the harmonic coefficients are of E and B type and the spectra contain the auto-, EE, BB, and cross-, EB, spectra.  As we treat the foreground as templates $\boldsymbol{\hat{a}}^{\rm{fg}}_j$ is not affected by the averaging over noise and CMB realisations and
 the matrix $\boldsymbol{\hat{a}}^{\rm{fg}}_j \boldsymbol{\hat{a}}^{\rm{fg,t}}_{j'}$ contains products of all multipole coefficients of polarised foreground component. Computing those may pose a significant challenge as they may depend on fine details of the foreground models. However, as discussed in~\cite{stompor2016} and shown below, in the computation of the likelihood we in fact only need the auto- and cross- spectra of all foreground signals. This not only speeds up the calculations but as the spectra are generally much better known, our predictions are more robust and reliable. \\
The first term of $\boldsymbol{E}_{jj'}$ depends explicitly on $\boldsymbol{W}^{\rm{cmb}}$
reflecting the effect of the estimation of the instrumental parameters in the generalised likelihood on the CMB content in the estimated CMB signal. As mentioned earlier, see~\cite{stompor2016}, in the case without instrumental effects, $ \boldsymbol{W}^{\rm{cmb}} \boldsymbol{A}^{\rm{cmb}} = \boldsymbol{1}$, the first term of $\boldsymbol{E}_{jj'}$ would reduce to $\boldsymbol{\mathcal{C}}^{\rm{cmb}}_\ell\delta_{jj'}$. The instrumental parameters also affect the second term of Eq.~\ref{eq:Ejj}. This term produces a non-vanishing contribution even in the absence of instrumental effects, it is however modified if they are present.

The last term concerns the noise power spectra $\boldsymbol{\mathcal{C}}^{\rm{noise}}_\ell$. This is the noise in the CMB map obtained after generalised component separation:
\begin{equation}
\begin{split}
    \boldsymbol{\mathcal{C}}^{\rm{noise}}_\ell &=  \boldsymbol{{W}}^{\rm{cmb}}  \frac{1}{2\ell+1}\sum_{m}\langle \boldsymbol{\hat{a}}^{\rm{noise}}_j \boldsymbol{\hat{a}}^{\rm{noise,t}}_{j} \rangle  \boldsymbol{{W}}^{\rm{cmb,t}} \\
    &= \left[\left(\boldsymbol{{\Lambda}}^t \boldsymbol{N}^{-1}_{\ell} \boldsymbol{{\Lambda}} \right)^{-1}\right]_{\rm{cmb} \times \rm{cmb}}
\end{split}
\end{equation}
where $\boldsymbol{N}_{\ell}$ for the frequency band $i$ is computed using the characteristics of the considered instrument, for instance for Simons Observatory it is given by the following formula with the SAT's $1/f$ power index \cite{Ade_2019}:
\begin{equation}
\label{eq:noise_ell}
    \boldsymbol{N}^{i}_{\ell} \equiv (w_i)^{-1} e^{ \left( \ell(\ell + 1) \frac{\text{FWHM}_i^2}{8\log{2}} \right)} \left( \left(\frac{\ell}{\ell_{knee}^i}\right)^{-2.4} + 1 \right)
\end{equation}
with $w_i^{-1}$ the sensitivity of the frequency channel $i$ in $(\mu K. \rm{rad})^2$. Here we also take into account the effect of the beam and inserted it by hand since, as mentioned in \cite{stompor2016}, for simplicity in the generalised spectral likelihood the noise is assumed to be white-like all the way down to the pixel scale and no beams are accounted for.
Here, $\text{FWHM}_i$ stands for the full-width half maximum for the $i^{\text{th}}$ channel, in radians. $1/f$ noise is also included via the last term and $\ell_{knee}^i$ is the position of the knee in harmonic space for the $i^{\text{th}}$ frequency band. We have assumed no frequency to frequency correlations here.

\subsubsection{Cosmological likelihood}

All the terms that compose our recovered CMB data in harmonic domain notwithstanding, in our cosmological likelihood we describe them as containing only the CMB signal and the noise. The corresponding covariance matrix, $\mathbf{C}$, is then given by,
\begin{equation}
\label{eq:cosmo_covariance}
    \boldsymbol{C}_{jj'} =
    \boldsymbol{\mathcal{C}}^{\rm{cmb, model}}_\ell (\theta)
    \delta_{jj'} + \boldsymbol{\mathcal{C}}^{\rm{noise}}_\ell\delta_{jj'}
\end{equation}
where the CMB covariance includes birefringence effect, described by the birefringence angle $\beta_b$, and the B-modes signal, described by the tensor-to-scalar ratio, $r$, and reads,
\begin{equation}
\label{eq:Cl_model}
\begin{split}
&\boldsymbol{\mathcal{C}}^{\rm{cmb, model}}_\ell (r, \beta_b) \\
&\equiv \boldsymbol{\mathcal{R}}(\beta_b)
\begin{pmatrix}
    C_\ell^{EE,\rm{p}} &   0 \\
   0 &r.C_\ell^{BB,\rm{p}} + A_L.C_\ell^{BB,\rm{lens}})
\end{pmatrix}
 \boldsymbol{\mathcal{R}}^{-1}(\beta_b)
\end{split}
\end{equation}
Here $C_\ell^{BB,\rm{p}}$ is a primordial B-mode spectrum computed for $r=1$ and $C_\ell^{BB,\rm{lens}}$ stands for the lensed B mode power spectrum. $A_L$ encodes the delensing where $A_L=1$ means no delensing and $A_L=0$ means total delensing. In the following we only consider $A_L=1$. $C_\ell^{EE,\rm{p}}$ is the E-mode spectrum including lensing and, as usual, the \emph{primordial} CMB EB cross spectra is set to zero. Models predicting non-zero primordial EB could be accounted for here in the covariance matrix, but we choose to ignore such models for simplicity and consider them in future work.
We ignore the effect of $r$ and delensing on the EE power spectrum.

The equation above assumes that isotropic birefringence acts on the lensed CMB spectra. This does not imply any loss of generality as isotropic birefringence commutes with lensing since it is scale independent. This is related to the fact that the lensing effect does not depend on the coordinate frame while the effect of isotropic birefringence can be seen as merely a coordinate change. The effect of a non-zero birefringence angle on the CMB power spectra is depicted in Fig~\ref{fig:spectra}.

We can then finally input our model and data in the cosmological likelihood which is computed case by case for each set of $\Gamma$ parameters drawn from the generalised spectral likelihood distribution, e.g.,~\textcite{Tegmark2008},
\begin{equation}
\label{eq:cosmo_like}
\begin{split}
     \langle S^{cos} \rangle &= \text{tr}\,\boldsymbol{C}^{-1}\boldsymbol{E} + \ln{\det{\boldsymbol{C}}} \\
     &= f_{sky}\sum_{\ell = \ell_{min}}^{l_{max}}{\frac{(2\ell+1)}{2}\left( Tr(\boldsymbol{C}^{-1}_{\ell}\,\boldsymbol{E}_{\ell}) + \ln(\det(\boldsymbol{C}_{\ell}))\right)},
\end{split}
\end{equation}
where,
\begin{equation}
    \boldsymbol{E}_\ell \, \equiv \, \frac{1}{2\ell+1}\,\sum_m\,\boldsymbol{E}_{jj}, \  \ \ \ \hbox{\textrm and}\; j = \ell^2+\ell+m.
\end{equation}
This shows that we only need to know $m$-averaged, diagonal elements of the data matrix, $\boldsymbol{E}$, which are fully defined by the cross-spectra of all the foreground signals as well as the CMB power spectra.
We note that this conclusion as well as Eq.~\ref{eq:cosmo_like} hold only if the noise is isotropic.

\subsection{Implementation}
\label{sub:algo}

The overall implementation of the method follows its main stages as described in the introduction of this section and as shown in Fig \ref{fig:method_diagram} in the case of the forecasting framework. Here, we provide a few more details concerning the implementation of some of the key stages for each of the two steps of the method.

On the first step, in order to evaluate spectral likelihood we need to estimate $\boldsymbol{S}^{\rm{cmb}}$, $\boldsymbol{F}$ (defined in Eq \ref{eq:S_CMB} and Eq \ref{eq:F_fg} respectively) and $\boldsymbol{N}$ for the CMB, foreground and noise signals. Those values are averaged over observed sky pixels as well, resulting in a $2\times2$ matrix encoding $QQ$, $UU$, and $QU$ correlations only. Here we are focusing on polarised signal only but it could be easily generalised to intensity signal as well. In practice to get $\boldsymbol{S}^{\rm{cmb}}$  we average 1000 CMB map realisations generated using the synfast function in \texttt{healpy}\footnote{\protect \url{http://healpix.sf.net}} \cite{Zonca2019,2005ApJ...622..759G} and using as input power spectra those generated by \texttt{CAMB}\footnote{\url{https://camb.info}} assuming the cosmological parameters estimated in the Planck 2018 release \cite{Planck2018_cosmo}, with $\ell_{\rm min} = 2$ and $\ell_{\rm max} = 4000$. In this paper we will use two sets of $(r,\beta_b)$ parameters that will be described in Section \ref{subsub:sim_CMB}. The assumed resolution, sky coverage, and the noise covariance, $\boldsymbol{N}$, depends on the instrument for which the forecast is performed.

The foregrounds covariance matrix, $\boldsymbol{F}$, is computed for noiseless frequency maps produced by \texttt{PySM} \cite{Thorne_2017} and subsequently averaged over all observed pixels. The experimental and cosmological parameters, as well as foreground models are detailed in section~\ref{sec:application}.

The component separation code used here is based on the ForeGroundBuster (\texttt{FGBuster}) library\footnote{\url{https://github.com/fgbuster/fgbuster}} and has been adapted to account for systematic effects such as polarisation angles and the addition of priors as mentioned earlier.
The sampling of the generalised spectral likelihood is performed using the \texttt{emcee} package~\cite{2013PASP..125..306F}. We used $2$ walkers per dimension, with $13,000$ steps and we burned the $5,000$ first ones. For the fiducial SO SAT-like case explored in this paper with $6$ miscalibration angles and $2$ foreground parameters it totals $128,000$ samples.

On the second step, in order to evaluate the cosmological likelihood we need to construct observed and model CMB power-spectra $\boldsymbol{\mathcal{C}}^{\rm{cmb}}_\ell$ we use the same Planck CMB power spectra templates as the one used for the generation of CMB maps in the first step and simulations.

We use the same frequency maps generated by \texttt{PySM} that were utilised in the first step to compute all the auto- and cross-spectra needed to compute $\boldsymbol{\hat{a}}^{\rm{fg}}_j \boldsymbol{\hat{a}}^{\rm{fg,t}}_{j'}$. Once computed for a given sky model and instrumental characteristics, the relevant $ \boldsymbol{{W}}^{\rm{cmb}} \boldsymbol{\hat{X}}$ factors are applied to have the contribution of foregrounds to the data after generalised component separation. It is only this last step that needs to be done for each sample of the spectral likelihood.
As in the first step, we are only using polarised power spectra EE, BB and EB but total intensity could easily be added if desired.

In order to propagate the statistical uncertainties incurred on the first step all the way to the estimation of cosmological parameters we perform double sampling in which for each sample (after burning) of the spectral likelihood we compute the corresponding CMB correlation matrix $\boldsymbol{E}_\ell$, Eq.~\ref{eq:Ejj}, and subsequently draw a sample from the corresponding cosmological likelihood Eq.~\ref{eq:cosmo_like}. To avoid any bias coming from initial values on each of these steps when drawing from the cosmological likelihood we use once again \texttt{emcee} with $300$ steps and a burn of $299$ so that we only keep the last point. Both samples put together constitute a single sample drawn from an effective joint distribution of spectral, instrumental, and cosmological parameters. We resort to this rather intricate way of sampling in order to alleviate biases due the method itself, first on the spectral/instrumental parameters and then, as a consequence, on the cosmological ones,~\cite{stompor2009}. The downside, in addition to the computational cost, is that the effective joint distribution we sample from is not merely a product of the spectral and cosmological likelihoods.

\section{Application}
\label{sec:application}
In this section we discuss the application of the forecasting method described in the previous section. We focus here on the case of a typical CMB ground-based telescope of third generation demonstrating the proposed framework and its performance on a specific experimental set-up as described in the next section. The framework is however general and can be applied to any other CMB experiment. Below we first describe the instrument configuration, followed by the sky simulation used in the application and finally the specific analysis assumptions that we consider in this work, such as the modelling of the instrumental response matrix $\boldsymbol{X}$.

\subsection{Instrument specifications}

For concreteness we use the configuration and noise specifications of the upcoming Simons Observatory's (SO) Small Aperture Telescopes (SAT) as described in~\cite{Ade_2019}.
The three SO SATs are planned to observe the sky in $6$ frequency channels: $27$, $39$, $93$, $145$, $225$ and $280$ GHz.
This will help to separate the CMB signal from the astrophysical foregrounds. \\
The SATs will observe $f_{\rm sky} \approx 10\%$ of the sky and generate sky maps with a typical resolution of $n_{\rm side}=512$ which corresponds to $\sim 6.8$ arcmin using the \texttt{HEALPix} convention~\cite{Zonca2019,2005ApJ...622..759G}. This results in around $3 \times 10^5$ observed sky pixels. Given the resolution, sky coverage and noise property we fix the multipole scales at $\ell_{\rm min}=30$ and $\ell_{\rm max} = 300$.\\
We use the publicly available code~\texttt{V3calc}\footnote{\url{https://github.com/simonsobs/so_noise_models}} and the sensitivities from the SO science goals and forecast paper~\cite{Ade_2019} to compute the sensitivity per frequency $w_i^-1$ after five years of observation for the high frequencies focal plane ($225$ and $280$ GHz), 5 and 4 years for two middle frequencies focal planes ($93$, $145$ GHz), and 1 year of observation for the low frequencies focal plane ($27$, $39$ GHz).
The resulting sensitivities of the baseline white noise case used are detailed in Table~\ref{tab:noiselvl}.
We then define the noise per sky pixel for each frequency band $i$, $\boldsymbol{N}_p^{i}$, used in Eq.~\ref{eq:ddt}, as,
\begin{equation}
    \boldsymbol{N}_p^{i} = \frac{ w_i^{-1}}{\Omega_{\rm{pix}}^2}
\end{equation}
where we converted the sensitivities from Table~\ref{tab:noiselvl} to $(\mu K. \rm{rad})^2$ to get  $w_i^{-1}$, as in Eq.\ref{eq:noise_ell}. And $\Omega_{\rm{pix}}^2$ is the area of a pixel in the sky in $\rm{rad}^2$.
Such noise covariance corresponds to the white noise, which is uncorrelated between pixels and frequencies. This turns out to be a reasonable assumption given the latest available SO simulations~\cite{BBpipe_paper} and sufficient for the spectral likelihood evaluation.\\
For the cosmological likelihood we take into account beam effects and $1/f$ noise in Eq.~\ref{eq:noise_ell}. For the computation of $\boldsymbol{N}^{i}_{\ell}$ we use the baseline sensitivity and so-called optimistic $1/f$ modes of SO SAT, as well as the SAT's beams as detailed in Table~\ref{tab:noiselvl}.

\begin{table}
    \centering
    \begin{tabular}{c||c|c|c|c|c|c}
        Frequency channel [GHz] & 27 & 39 & 93 & 145 & 225 & 280 \\
        \hline
        Polarisation sensitivity [$\mu$K-arcmin] &
        49 & 30 &  3.8 & 4.7 & 9.0 & 23 \\
        \hline
        $\ell_{knee}$ & 15 & 15 & 25 & 25 & 35 & 40 \\
        \hline
        FWHM [arcmin] & 91 & 63 & 30 & 17 & 11 & 9
    \end{tabular}
    \caption{SO SAT baseline white noise levels,
    $1/f$ noise properties,
    and FWHMs.}
    \label{tab:noiselvl}
\end{table}
For the calibration priors we typically assume  a fiducial precision of $\sigma(\alpha) = 0.1^\circ$. This is on the conservative side for a drone-borne calibrator currently validated on several telescopes in the Atacama such as ACT \cite{Fowler:07,Thornton2016} and CLASS \cite{Essinger_Hileman_2014} and also planned to be applied to SO SATs. Current forecast for this method is $0.01^\circ\leq \sigma(\alpha) \leq 0.1^\circ$, which should be achievable in several frequency channels \cite{Nati_2017}. Other approaches, such as a mobile rotating wire grid or astrophysical sources give typically comparable but somewhat worse precision, $< 1^\circ$ for the grid~\cite{Bryan_2018} and $\simgt 0.3^\circ$ for Tau A~\cite{Aumont_2020, Ritacco_2018}. Therefore in the following we discuss the impact of the calibration precision on our conclusions.

\subsection{Input Sky Simulations}
\subsubsection{Foregrounds} \label{subsub:input_fg}

As the fiducial foreground case we take the ``d0s0'' model of~\texttt{PySM}~\cite{Thorne_2017}. For dust, this assumes a modified black-body parameterised by its spatially-constant temperature $T_d$ and spectral index $\beta_d$.
The synchrotron emission is modelled as a power law characterised by the constant spectral index $\beta_s$. The spectral emission densities for those two components are expressed as:
\begin{eqnarray}
    \centering
    Y_{\rm{sync},p} (\nu, \beta_s) &=& Y_{\rm{sync},p}(\nu_{0,s}) \left(\frac{\nu}{\nu_{0,s}}\right)^{\beta_s}, \\
    Y_{\rm{dust},p} (\nu, \beta_d, T_d) &=& Y_{\rm{dust},p}(\nu_{0,d}) B(\nu,T_d) \left(\frac{\nu}{\nu_{0,d}}\right)^{\beta_d},
    \label{eqn:scalings}
\end{eqnarray}
where $Y$ can be a $Q$ or $U$ map of a component, expressed in ${\rm MJy/sr}$. $Y_{\rm{sync},p}(\nu_{0,s})$ and $Y_{\rm{dust},p}(\nu_{0,d})$ are the template maps for synchrotron and dust at their respective reference frequencies $\nu_{0,s} = 23$ GHz and $\nu_{0,d} = 353$ GHz that are then scaled at the frequencies of interest for observations. $B(\nu,T_d)$ is a black body at temperature $T_d$ and frequency $\nu$. In the \texttt{PySM} ``d0s0'' simulation, the spatially constant spectral parameters are based on Planck results \cite{Planck2015_compsep} and are given by:
\begin{equation}
    \beta_d = 1.54, \quad T_d = 20{\rm K}, \quad \beta_s = -3.
    \label{eq:d0s0_values}
\end{equation}
As examples of a more complex foreground model, we use the ``d1s1'' model which is similar to``d0s0'' but allows for spatial variability of the spectral parameters,  $\beta_d(p)$, $T_d(p)$ and $\beta_s(p)$, as well as the model referred to as
``d7s3''. Here ``s3'' denotes the synchrotron model which adds an extra curvature term, $\mathcal{C}$, to the standard power law frequency scaling, which is constant over the sky:
\begin{equation}
        Y_{\rm{sync},p} (\nu, \beta_s) = Y_{\rm{sync},p}(\nu_{0,s}) \left(\frac{\nu}{\nu_{0,s}}\right)^{\beta_s(p) + \mathcal{C}\ln(\nu / \nu_0)}
\end{equation}
The dust model ``d7'' uses as a template the same Planck $353$ GHz map as the other two models but the frequency scaling used is based on dust grain models with different physical properties, shape, size, temperature described in detail in \cite{Draine_2013,hensley_thesis}. This dust model does not have an analytic function to describe the frequency scaling of the dust template and is therefore a good benchmark to test the parametric component separation since it does not trivially conform with the assumptions of the method.

\subsubsection{CMB} \label{subsub:sim_CMB}
For the CMB we set all the parameters to the best-fit values provided by Planck \cite{Planck2018_cosmo} and only vary two parameters, the tensor-to-scalar ratio $r$ and the birefringence angle $\beta_b$. We do not consider delensing in this work, i.e. $A_L = 1$ in Eq.~\ref{eq:Cl_model}.
However this can be straightforwardly included in our framework. The input CMB power spectra used to get the frequency maps (as described in the previous section) are computed using the same equation as the model CMB in Eq.~\ref{eq:Cl_model}. We choose two sets of cosmological parameters in addition to the fiducial $\Lambda$CDM cosmological parameters from Planck 2018 \cite{Planck2018_cosmo}:
\begin{eqnarray}
    \centering
    r &=& 0.0, \quad \beta_b = 0.0^{\circ}\\
    r &=& 0.01, \quad \beta_b = 0.35^{\circ}
\end{eqnarray}
With $\beta_b = 0.35^{\circ}$ corresponding to the central value found in the work of \textcite{Minami_2020}.
\\

\subsubsection{Instrumental Effects}
In the simulation we consider the effect of a potential miscalibration of the polarisation angle of each of the frequency bands of the telescope.
We model this effect assuming that each recovered single frequency map has its own independent polarisation angle. This angle is then to be understood as an effective angle resulting from detector-level miscalibration averaged during the map-making procedure.
This assumption can be adjusted as needed: our approach is generalisable to a miscalibration angle per focal plane, one per wafer, one per pixels, etc. We also assume that the miscalibration angle is the same for all pixels of the considered maps. Again this can be relaxed if needed.
In this work, for concreteness, we assume some specific, true values of the miscalibration angle for each map. They are summarised in Table~\ref{tab:input_miscal}. But the presented results do not depend on the specific values assumed.
\renewcommand{\arraystretch}{1.4}

\begin{table}[h]
    \centering
    \begin{tabular}{c|c|c|c|c|c|c}
        Frequency channel [GHz] & 27 & 39 & 93 & 145 & 225 & 280 \\
        \hline
        Input polarisation angle [$^\circ$] & 1 & 1.66 & 2.33 & 3 & 3.66 & 4.33
    \end{tabular}
    \caption{Input polarisation angle per frequency bands}
    \label{tab:input_miscal}
\end{table}

\subsection{Analysis Model and Priors} \label{sub:analysis_hyp}

For the analysis we assume the foreground scaling model as in Eqs~\ref{eqn:scalings} with both dust and synchrotron parameters assumed constant over the sky. Moreover, we fix dust temperature to $T_d = 20$K as suggested by Planck results~\cite{Planck2015_compsep} given that the SO SATs do not have enough high frequency observation bands to discriminate between $T_d$ and $\beta_d$. Importantly, this is in agreement with one of the ``d0s0'' assumption, Eq.~\ref{eq:d0s0_values} and therefore,
the foreground model assumed in the analysis allows for an accurate description of the data simulated in the case of the ``d0s0'' model of PySM. As it is discussed this is however not the case for the ``d1s1'' and ``d7s3'' models. For the instrumental matrix $\boldsymbol{X}$ we assume an isotropic rotation of the polarisation angle for each of the frequencies, exactly as for the sky simulations with rotations angles at each frequency treated as free parameters.
We use Gaussian priors with a precision of $\sigma_{\alpha_i} = 0.1^\circ$, Eq.~\ref{eq:spec_like_prior},  in the spectral likelihood to break the degeneracies between these parameters.
Unless specified otherwise the priors are centred on the true input polarisation angles.

\section{Results}
\label{sec:results}

First we present the results of the first analysis step that retrieves the spectral indices and the miscalibration angles from noisy, foreground-contaminated and miscalibrated frequency maps. Then we present the results of the second step that constrains cosmological parameters. We discuss various examples to demonstrate the effects of the different contaminants on the cosmological parameters estimation, such as biases or precision loss. We then investigate the dependence of the constraint on cosmological parameters as a function of prior precision. And finally we explore the case where calibration priors are biased and their impact on instrumental and cosmological parameters estimations. \\

\begin{figure*}[t]
\centering
\includegraphics[width=0.7\paperwidth]{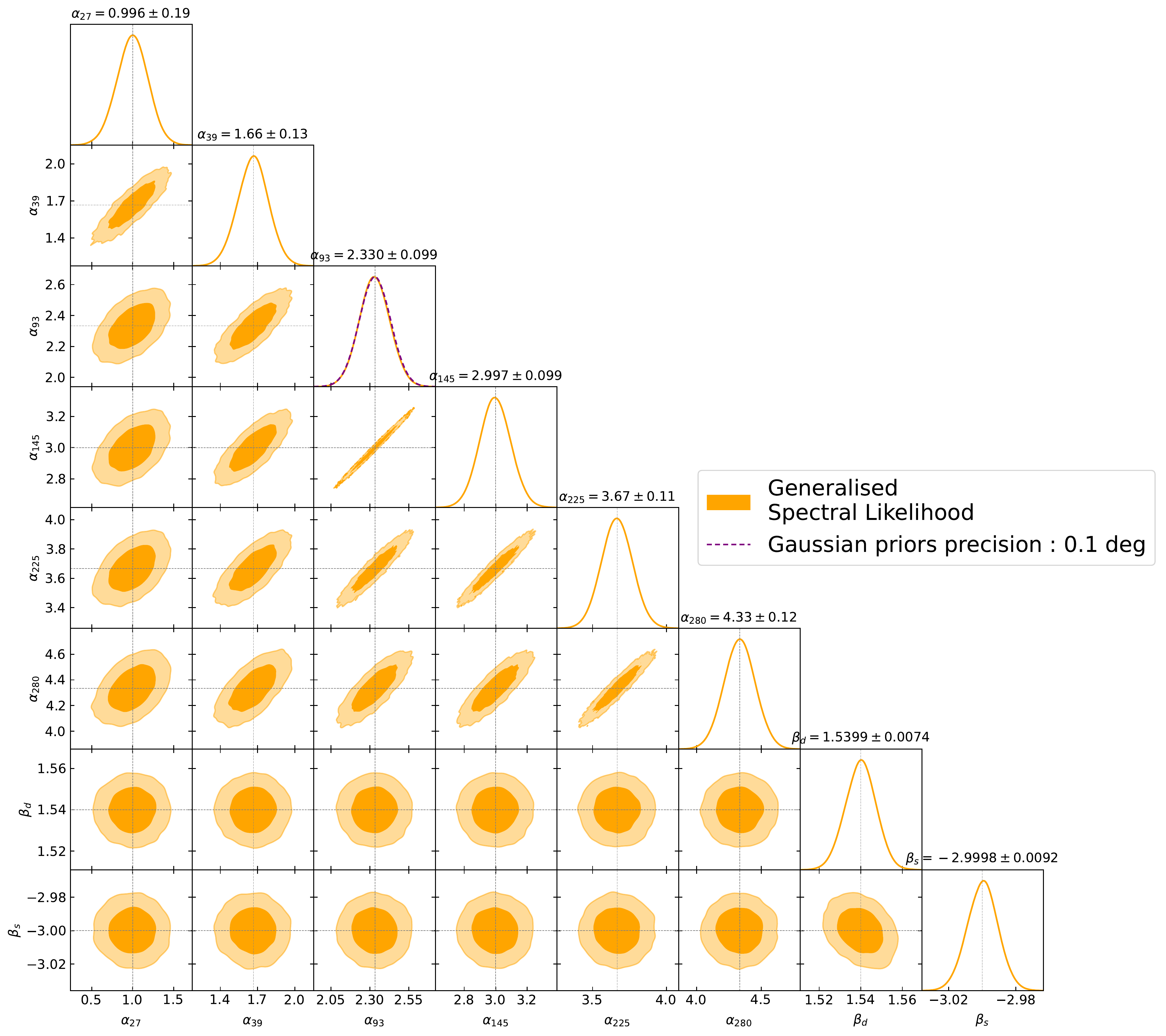}
\caption{Results of the generalised spectral likelihood with ``d0s0'' input foregrounds model. There is only one prior on the $93$ GHz channel, with a precision of $\sigma_{\alpha_{93}} = 0.1^\circ$. The dashed purple lines correspond to the Gaussian priors. The orange contours correspond to the sampling of the generalised spectral likelihood. The grey dotted lines correspond to the input values.}
\label{fig:spectral_s0d0_1prior}
\end{figure*}

\subsection{Fiducial Case: ``d0s0'', $\boldsymbol{r=0}$, $\boldsymbol{\beta_b = 0^\circ}$}
We start with the simplest case of the ``d0s0'' \texttt{PySM} model that assumes constant foreground spectral parameters across the sky.
Since consistent assumptions were used to build the mixing matrix $\boldsymbol{A}$, Eq.~\ref{eq:data_model} --- which is used throughout this work --- should accurately describe the input foreground maps.
\\
We explore two cases, one with a prior on each frequency channel and the other with a prior on the $93$ GHz channel only. In both cases the prior precision is $\sigma_{\alpha_i} = 0.1^\circ$.

\renewcommand{\arraystretch}{1.2}

\begin{table*}
    \centering
\begin{tabular*}{0.925\textwidth}{l|c|c||c|c||c||c}
       \hline
        Foreground input & \multicolumn{2}{c||}{``d0s0''} & \multicolumn{2}{c||}{``d1s1''} & \multicolumn{1}{c||}{``d7s3''} & \multicolumn{1}{c}{``d0s0'' $\star$}\\
        \hline
        Number of priors& \multicolumn{1}{c|}{1} & \multicolumn{1}{c||}{6} & \multicolumn{1}{c|}{1} & \multicolumn{1}{c||}{6} & \multicolumn{1}{c||}{6} &  \multicolumn{1}{c}{6} \\
\hline
  $\alpha_{27}[^\circ]$  & $1.0\pm 0.2$              & $1.00\pm 0.08$             & $1.0\pm 0.2$               & $1.01\pm 0.08$             & $1.01\pm 0.08$             & $1.00\pm 0.08$             \\
 $\alpha_{39}[^\circ]$  & $1.7\pm 0.1$               & $1.67\pm 0.05$             & $1.7\pm 0.1$               & $1.67\pm 0.05$             & $1.67\pm 0.05$             & $1.67\pm 0.05$             \\
 $\alpha_{93}[^\circ]$  & $2.3\pm 0.1$             & $2.33\pm 0.05$             & $2.3\pm 0.1$               & $2.33\pm 0.05$             & $2.33\pm 0.05$             & $2.33\pm 0.05$             \\
 $\alpha_{145}[^\circ]$ & $3.0\pm 0.1$             & $3.00\pm 0.05$             & $3.0\pm 0.1$               & $3.00\pm 0.05$             & $3.00\pm 0.05$             & $3.00\pm 0.05$             \\
 $\alpha_{225}[^\circ]$ & $3.7\pm 0.1$               & $3.67\pm 0.05$             & $3.7\pm 0.1$               & $3.66\pm 0.05$             & $3.66\pm 0.05$             & $3.66\pm 0.05$             \\
 $\alpha_{280}[^\circ]$ & $4.3\pm 0.1$               & $4.33\pm 0.06$             & $4.3\pm 0.1$               & $4.33\pm 0.06$             & $4.33\pm 0.06$             & $4.33\pm 0.06$             \\
 $\beta_d$      & $1.540\pm 0.007$           & $1.540\pm 0.007$           & $1.575\pm 0.008$           & $1.575\pm 0.007$           & $1.377\pm 0.007$           & $1.540\pm 0.007$           \\
 $\beta_s$      & $-3.000\pm 0.009$          & $-3.000\pm 0.009$          & $-3.006\pm 0.009$          & $-3.006\pm 0.009$          & $-3.046\pm 0.009$          & $-3.000\pm 0.009$          \\

 \hline
 \hline
 $r$                    & $0.000\pm 0.002$             & $0.000\pm 0.002$           & $0.002\pm 0.002$           & $0.002\pm 0.002$           & $0.002\pm 0.002$           & $0.010\pm 0.002$ \\
 ${\beta_b}[^\circ]$    & $0.0\pm 0.1$               & $0.00\pm 0.07$             & $0.0\pm 0.1$               & $0.00\pm 0.07$             & $0.00\pm 0.07$             & $0.35\pm 0.07$             \\
\hline
\end{tabular*}
    \caption{Summary of results for different input foreground models and instrumental parameters. All the priors used here have the precision $\sigma_{\alpha_i}=0.1^\circ$. $\star$ In the last column (``d0s0'') the input cosmological parameters are $r=0.01$ and $\beta_b=0.35^\circ$.}
    \label{tab:result_summary}
\end{table*}
\subsubsection{\textbf{Prior on a Single Channel}} \label{subsub:1prior}
First we consider a prior on the $93$ GHz channel only with a precision of $\sigma_{\alpha_{93}}=0.1^\circ$. We choose this channel as at this frequency the foregrounds amplitude is close to minimal as compared to the CMB signal and this is where most of the calibration effort is currently being allocated. \\
\paragraph{\textbf{Generalised Spectral Likelihood Results:}}
As described in Section~\ref{sub:algo} we explore the generalised spectral likelihood, Eq.~\ref{eq:spec_like_prior} with help of MCMC sampling.
The results are shown in Fig.~\ref{fig:spectral_s0d0_1prior} where the orange contours are obtained from the MCMC samples, the purple dashed line represents the Gaussian prior on the $93$ GHz channel and the grey dashed lines the input parameters. The $1\sigma$ statistical errors of the parameter estimations are detailed in Table~\ref{tab:result_summary}.\\
We notice that with only one prior on one polarisation angle we are able to have an unbiased estimate for all $6$ polarisation angles and $2$ spectral indices. Indeed the fact that we use all 6 frequency maps simultaneously in the generalised spectral likelihood
allows for deriving tight constraints on the relative angles of all the considered frequency channels with respect to a global ``instrument'' orientation angle.
The role of the prior is then to constrain the global angle which is necessary and sufficient to break the likelihood degeneracy.

The accuracy with which we can estimate the absolute polarisation angles for any of the channels is therefore limited by the prior precision as summarised in Table~\ref{tab:result_summary}. The $93$ GHz channel achieves the best precision on polarisation angle, $\sigma(\alpha_{93})=0.099^\circ$, which corresponds to the prior precision (within the accuracy provided by the sampling). All other channels show larger errors as they include the error on the relative angle as set by the likelihood problem. The overall increase of the error is subdominant as compared to the prior-driven constraint on the global angle, showing that indeed the relative angles are constrained with high precision.\\
Furthermore we are able to retrieve the foreground spectral parameters with a precision comparable
with the standard version of parametric component separation using \texttt{FGBuster} applied to a SO-like case but without instrumental parameter estimation~\cite{Ade_2019}.

As expected, the generalised spectral likelihood yields unbiased estimates of instrumental and spectral parameters, which we then use on the next step: the estimation of cosmological parameters.

\begin{figure}
\centering
\includegraphics[width=\linewidth]{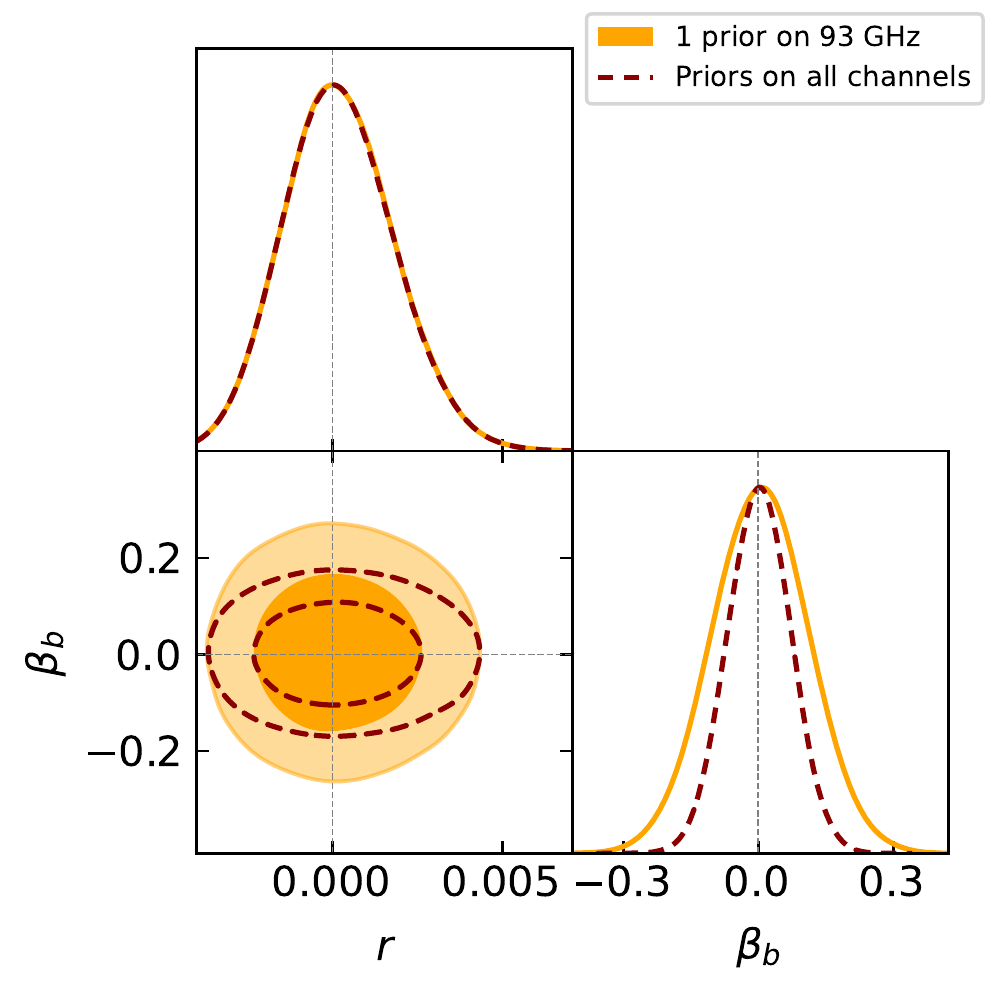}
\caption{Cosmological likelihood sampling, Eq.~\ref{eq:cosmo_like}, after foreground cleaning and systematic effect correction with \texttt{PySM} \emph{``d0s0''} as input and only \emph{one} prior on the $93$ GHz polarisation angle (orange). Dashed dark-red contours correspond to the case with priors on all polarisation angles. The grey dashed lines correspond to the input values. The central values and error bars are in Table ~\ref{tab:result_summary}.}
\label{fig:cosmo_d0s0}
\end{figure}

\paragraph{\textbf{Cosmological Likelihood Results:}}
As detailed in section~\ref{sub:algo}, for each sample of the spectral likelihood displayed in Fig.~\ref{fig:spectral_s0d0_1prior} we draw one sample of the corresponding cosmological likelihood.
This approach allows us to efficiently sample the full distribution, without reintroducing any method-related biases. The result is shown in Fig.~\ref{fig:cosmo_d0s0} where the orange contours are the MCMC samples obtained assuming a single $93$ GHz prior, and the grey dashed lines the input parameters. The estimations of both $r$ and $\beta_b$ are unbiased and with $r=0.0002^{+0.0015}_{-0.0017}$ which is compatible with the SO SATs published forecasts~\cite{Ade_2019}, and $\sigma(\beta_b) = 0.11^\circ$ which is $10\%$ bigger than the error bars expected from prior precision alone, which can be explained by the presence of noise and cosmic variance as we will see in Section \ref{sub:cosmo_wrt_prior}.

We conclude that in the case with simple foreground SEDs that match our model, and with non-zero polarisation angles, the method leads to unbiased estimates of spectral and hardware parameters, and provides competitive results on $r$ and an estimation of the birefringence angle $\beta_b$ limited only by the prior precision.

\begin{figure*}[t]
\centering
\includegraphics[width=0.7\paperwidth]{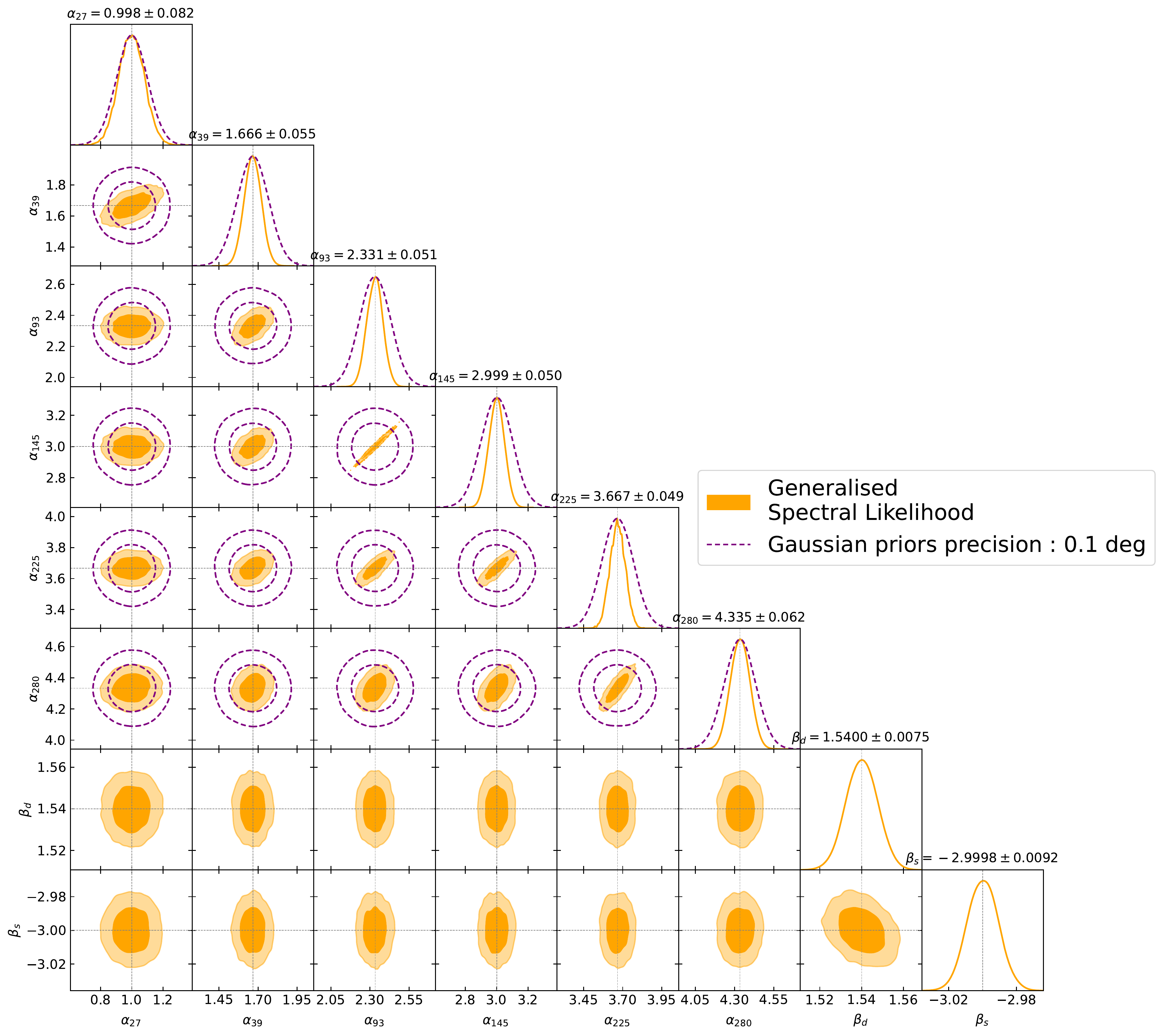}
\caption{Results of the generalised spectral likelihood with ``d0s0'' as input foregrounds model. The priors are on all 6 polarisation angles, with a precision of $\sigma_{\alpha_i} = 0.1^\circ$. The dashed purple lines correspond to the Gaussian priors. The orange contours correspond to the sampling of the generalised spectral likelihood. The grey dotted lines correspond to the input values.}
\label{fig:spectral_fid}
\end{figure*}

\subsubsection{\textbf{Priors on All 6 Channels}} \label{subsub:fid6prior}
We now investigate the case with priors on all 6 polarisation angles. We proceed in a similar fashion as in the previous case.
\paragraph{\textbf{Generalised Spectral Likelihood Results:}}
Fig.~\ref{fig:spectral_fid} shows the results of the MCMC sampling of the generalised spectral likelihood. Comparing the priors (dashed purple) and the samples (orange) we see that, contrary to the previous case,
the precision of the polarisation angle estimation is better than the prior precision, $\sigma_{\alpha_{i}}=0.1^\circ$, assumed for all frequency channel. This is consistent with the fact that the likelihood itself sets tight constraints on the relative angles for each frequency channel. So while the priors concern different objects, polarisation angles for their respective frequency channel, each of them effectively constrains the very same global polarisation angle. We thus expect that the actual constraint on this angle goes down roughly as one over square root of the number of frequency channels (however as the precision of priors gets better other contributions to the error bar, such as noise, become dominant as in this case, we explore this in more details in paragraph~\ref{sub:cosmo_wrt_prior}).  As the global angle uncertainty constitutes the  biggest contribution to the uncertainty of polarisation angle for each channel, we expect that the errors on these angles also decrease with the number of frequency channels in roughly the same way. \\

\paragraph{\textbf{Cosmological Likelihood Results:}}
The distribution of cosmological parameters after the generalised component separation with 6 priors is presented as dashed dark-red curves in Fig.~\ref{fig:cosmo_d0s0}. Again, the estimations of both $r$ and $\beta_b$ are unbiased and with a precision of $\sigma(r)\approx 2 \times 10^{-3}$
and $\sigma(\beta_b) \approx 0.07^\circ$ as mentioned in Table~\ref{tab:result_summary}. The estimation of $r$ is therefore unchanged with respect to the previous case but the estimation of $\beta_b$ has improved as a consequence of the improvement of the polarisation angle estimation in the first step. Having polarisation angle calibration on multiple frequency bands would therefore improve $\sigma(\beta_b)$ without necessarily requiring a large improvement of the calibration precision itself which can be very challenging.

\subsection{Complex Foregrounds, $\boldsymbol{r=0}$, $\boldsymbol{\beta_b = 0^\circ}$}

\begin{figure*}
\centering
\includegraphics[width=0.7\paperwidth]{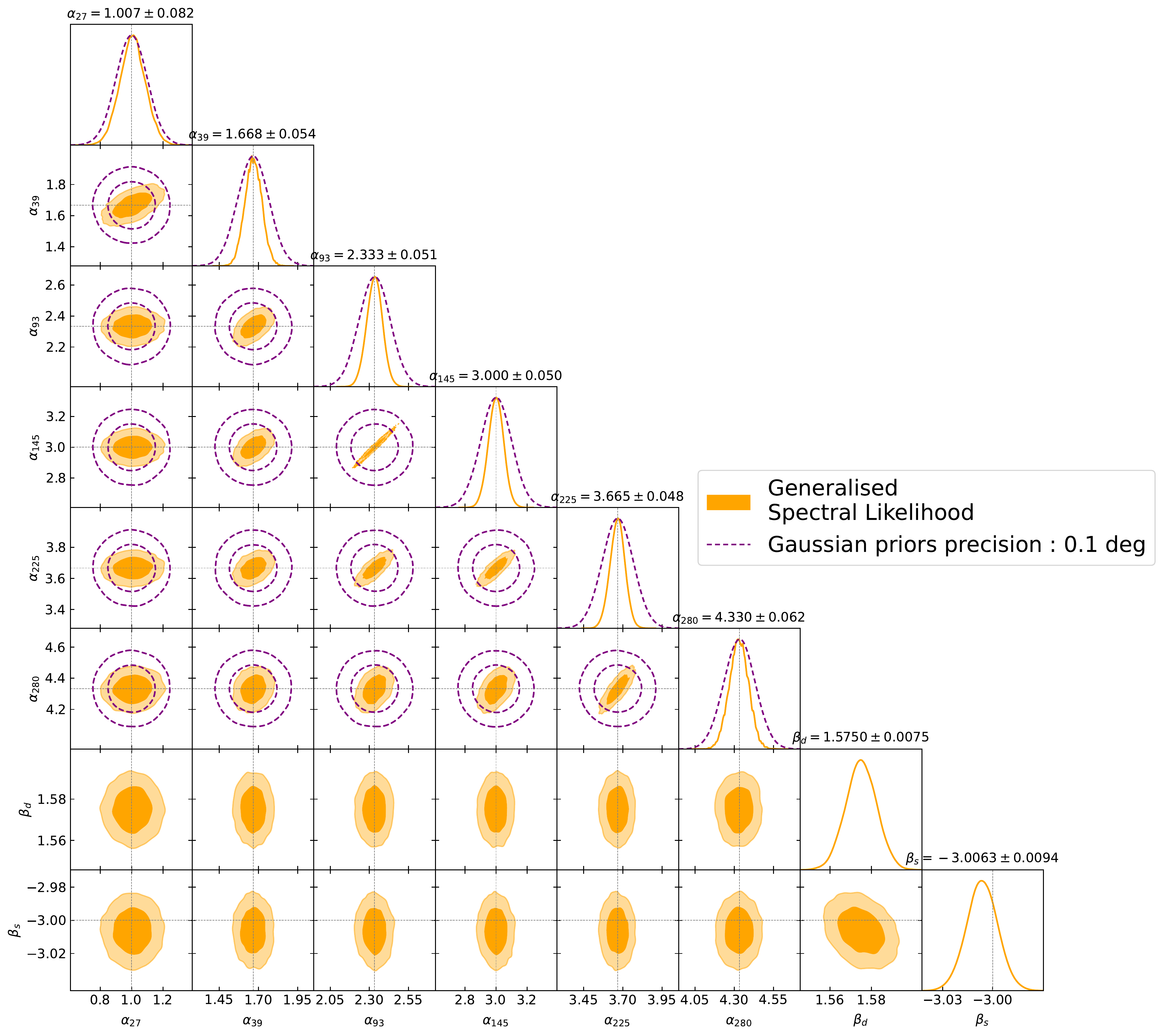}
\caption{Results of the generalised spectral likelihood with \textbf{d1s1} as input foregrounds model with the calibration priors on all 6 polarisation angles, with a precision of $\sigma_{\alpha_i} = 0.1^\circ$. The dashed purple lines show these Gaussian priors, while the orange contours correspond to the sampling of the generalised spectral likelihood. The grey dotted lines mark the input values.}
\label{fig:spectral_d1s1_6priors}
\end{figure*}

\subsubsection{Spatially-Varying Foreground SEDs: ``d1s1''}
This foreground model used in the simulated \emph{data} implements spatially-varying spectral indices.
However, the model we use to describe the data still assumes constant spectral indices. This may potentially lead to bias on cosmological parameters induced by the mismatch between foreground model and data.
For conciseness we focus on the case with \emph{priors on all polarisation angles}. The results obtained with one prior on the $93$ GHz channel are detailed in Table~\ref{tab:result_summary}.

\paragraph{\textbf{Generalised Spectral Likelihood Results:}}
Fig.~\ref{fig:spectral_d1s1_6priors} shows the results of the generalised spectral likelihood sampling. The estimation of polarisation angles are not significantly affected by the more complex foregrounds and results are similar to the previous, $6$-priors case. The spectral likelihood still manages to estimate  effective values of spectral indices, even if in the simulated data they vary from pixel to pixel.

\begin{figure}
\centering
\includegraphics[width=\linewidth]{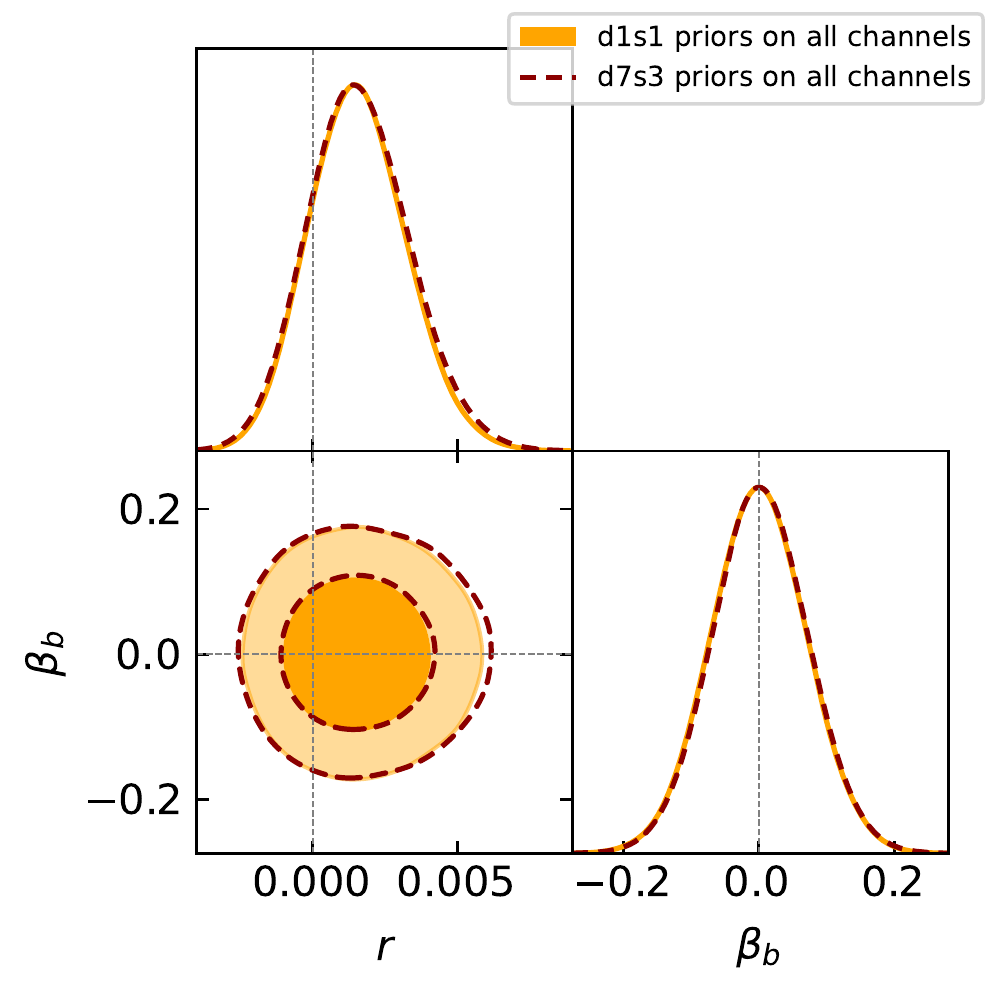}
\caption{Cosmological likelihood sampling after foreground cleaning and systematic effect correction with priors on \emph{all} polarisation angles. Results obtained for \emph{d1s1} (resp. \emph{d7s3}) are shown as orange contours (resp. dashed dark-red). The grey dashed lines correspond to the input values.}
\label{fig:cosmo_d1s1_6priors}
\end{figure}

\paragraph{\textbf{Cosmological Likelihood Results:}} Results are shown in Fig.~\ref{fig:cosmo_d1s1_6priors}. The estimation of the birefringence angle $\beta_b$ is not significantly affected by the complex foregrounds and by the mismatch between data and model.
This seems consistent with the unbiased estimate of polarisation angles in the previous step, and the foreground leakage to the recovered CMB EB correlation seems under-control.
For $r$ the estimation is slightly biased with $r\sim 0.0016$ but it is still $1\sigma$ compatible with the input, $r=0$ value.
The relatively small effects of foreground SEDs mismatch is mostly thanks to the small sky fraction observed by SO SATs ($f_{sky}=0.1$), their limited frequency coverage and their large angular scale, $\ell_{\rm min}=30$. Spatially varying spectral indices on e.g. a larger sky patch or with a larger frequency coverage would certainly bias more significantly both parameters.

\subsubsection{Non-Parametric Dust Model and Curved Synchrotron: ``d7s3''}
As described in \ref{subsub:input_fg} the ``d7'' dust model is non-parametric and should therefore pose particular problems to our generalised parametric component separation based on the simplest, pixel-independent scaling relations. The ``s3'' synchrotron model adds complexity as well with a global curvature term not taken into account in our data model.
Nevertheless we find that polarisation angles do not seem to be affected by non-parametric input foregrounds as can be seen in Table~\ref{tab:result_summary}.
For the cosmological likelihood, once again the limited sky fraction used by SO SATs helps to alleviate the impact of spatial variations and the generalised spectral likelihood sampling gives similar results as in the ``d1s1'' case. The impact on the cosmological likelihood is therefore limited as well, as shown in Fig.~\ref{fig:cosmo_d1s1_6priors} as dashed dark-red contours. The error bars are slightly different as described in Table~\ref{tab:result_summary} with a $\sim 5\%$ increase in the upper error bar on $r$ with respect to the ``d1s1'' case, and a $\sim 1.4\%$ \textit{decrease} on $\sigma(\beta_b)$, and with no detectable biases. \\

\subsection{Non-zero Cosmological Parameters: $\boldsymbol{r=0.01}$, $\boldsymbol{\beta_b = 0.35^\circ}$.}
For completeness we now focus on a case where the two cosmological parameters considered here are non-zero. We also use the input foreground model ``d0s0'' and priors on all channels. We do not show the results of the generalised spectral likelihood as they are essentially identical to the ones presented in Fig.~\ref{fig:spectral_fid}.

\begin{figure}
\centering
\includegraphics[width=\linewidth]{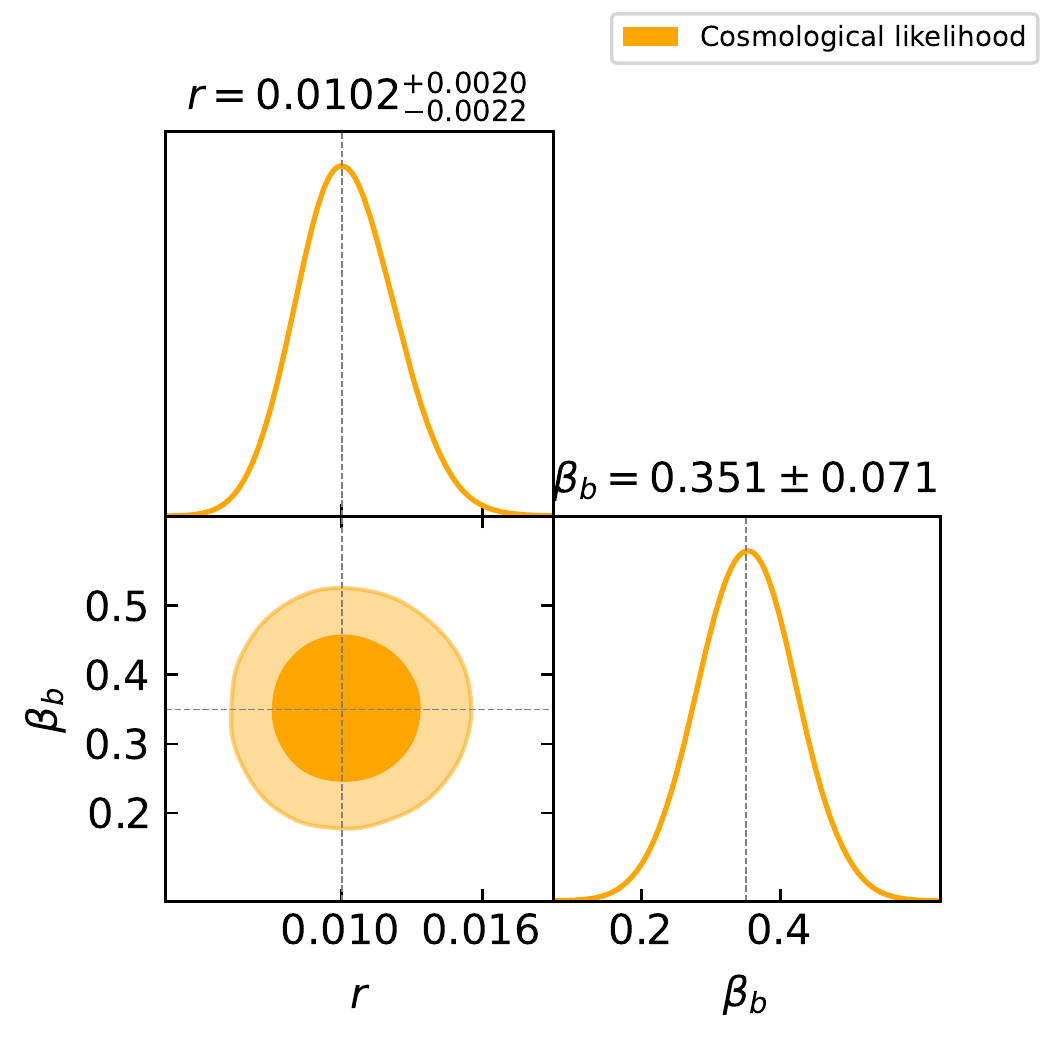}
\caption{Cosmological likelihood sampling after foreground cleaning and systematic effect correction with ``d0s0'' as foreground input and priors on \emph{all} polarisation angles. Input cosmological parameters are $r=0.01$ and $\beta_b=0.35^\circ$. The grey dashed lines correspond to the input values. This figure is analogous to  Fig.~\ref{fig:cosmo_d0s0} but assumes a non-zero birefringence angle.}
\label{fig:cosmo_d0s0_6priors_rb!=0}
\end{figure}

Fig.~\ref{fig:cosmo_d0s0_6priors_rb!=0} displays the cosmological constraints and indicates the possibility of detecting $r=0.01$ with a $\sim5\sigma$ precision, consistent with previous forecast \cite{Ade_2019}. A $\sim 5 \sigma$ detection of the value of $\beta_b=0.35^\circ$ as derived recently from the Planck data \cite{Minami_2020,Diego_Palazuelos_2022} seems achievable as well.

The results obtained for the other foreground models,``d1s1'', ``d7s3'', or with only a single prior are analogous to the corresponding cases with $r=0$ and $\beta_b=0^\circ$ and we do not show the likelihood plots here again.
In particular, we find that the results on $r$ are becoming progressively more biased but the biases never exceed $1 \sigma$ error bars. The results on $\beta_b$ are consistently unbiased with error bars going from $0.1^\circ$ to $0.07^\circ$ for the 1 and 6 priors cases respectively.

\subsection{Cosmological Parameters Estimation as a Function of Prior Precision}
\label{sub:cosmo_wrt_prior}
To better understand the dependence of the results on the prior precision and on the number of priors we perform the previous analyses with several values for the calibration precision. Conversely, this also provides specification for the calibration campaigns given pre-defined science goals.

\begin{figure}
\centering
\includegraphics[width=\linewidth]{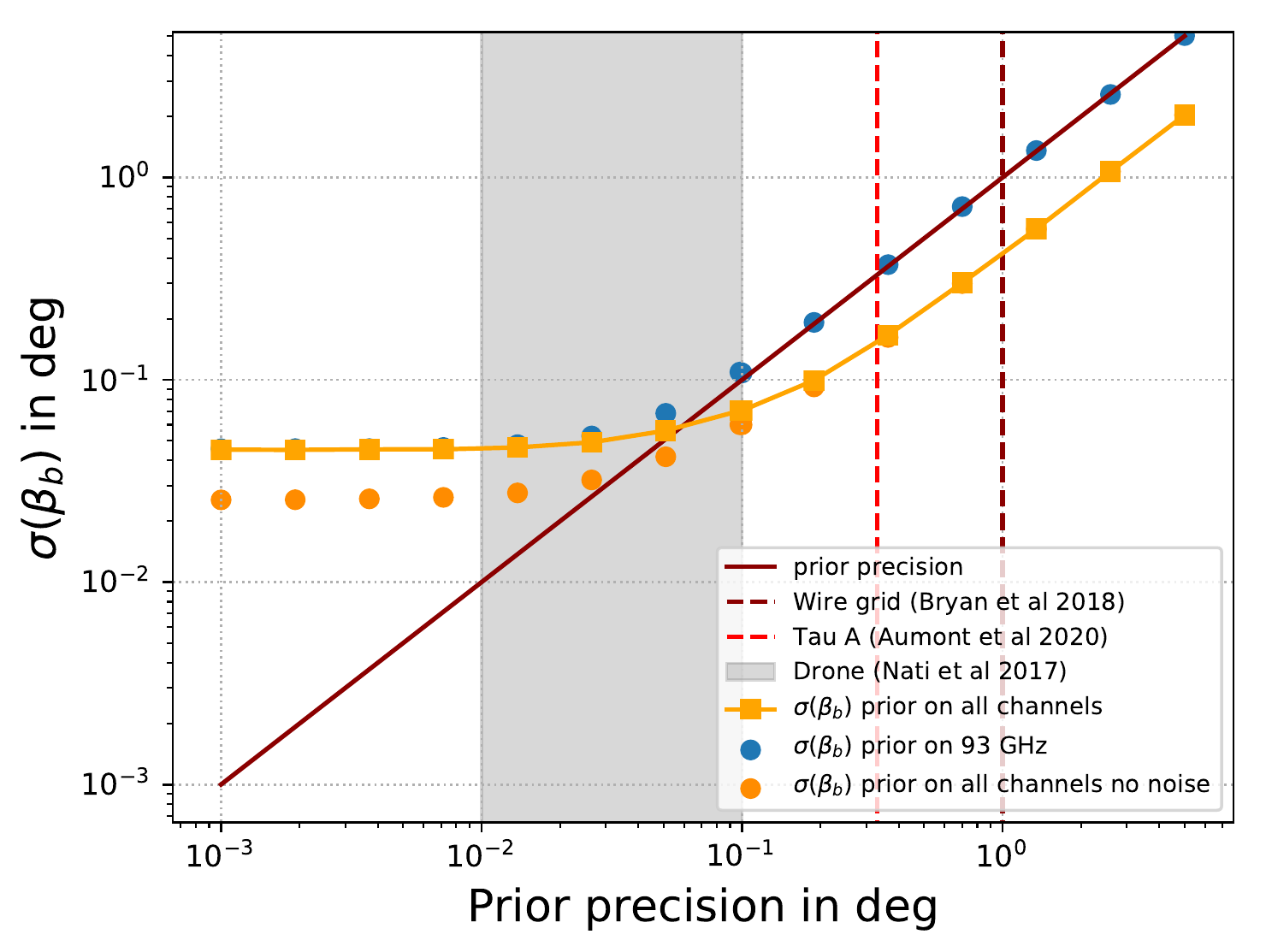}
\caption{Dependence of $\sigma(\beta_b)$ on prior precision for different calibration and noise scenarios as discussed in Section~\ref{sub:cosmo_wrt_prior}.}
\label{fig:sigma_beta}
\end{figure}

We limit ourselves to the case where $r=0$, $\beta_b=0^\circ$, and use the simple ``d0s0'' foregrounds in the input data. We allow the prior precision to change in the range  $0.001^{\circ}\leq\sigma_{\alpha_{i}}\leq5^{\circ}$ and consider cases with a single prior on the $93$ GHz channel and with priors on all polarisation angles for all channels. The dependence of $\sigma(\beta_b)$ on $\sigma_{\alpha_{i}}$ is displayed in Fig.~\ref{fig:sigma_beta}. The blue points correspond to the single prior case: it is clear that for large values of $\sigma(\alpha_i) \simgt 0.05^\circ$ the obtained precision on $\beta_b$ is determined by the prior precision. For smaller values of $\sigma_{\alpha_{i}}\lesssim 0.05^\circ$
the precision on $\beta_b$ saturates and reaches a plateau at $\sigma(\beta_b)\approx0.045^\circ$. This plateau is due to the cosmic and noise variances which dominate the error budget over the prior precision.
A similar behaviour is observed in the $6$ priors case. However, for large $\sigma_{\alpha_{i}}$, the obtained values if $\sigma(\beta_b)$ are now roughly a factor of $\sqrt{\#prior}$ lower than in the single prior case. As discussed already earlier in section~\ref{subsub:fid6prior}, this is because each prior corresponds effectively to an independent measurement of the global polarisation angle and the effective error on it therefore decreases with the number of channels. As the error is smaller for large prior uncertainties compared to the one prior case the dependence starts reaching the plateau somewhat earlier (i.e., for larger values of the prior precision) as its level remains the same whatever is the number of priors.

This result can help us with future calibration requirements and suggests that as long as we are in the prior dominated regime, to achieve a given $\sigma(\beta_b)$ one must either improve the precision of the calibration method, or up to a certain limit depending on the number of channels, calibrate several frequency channels to get the same results. As improving the absolute precision of calibration is quite challenging, multiplying calibration campaigns to other frequency channels seems to be a reasonable option. \\
Furthermore, to see how both noise and cosmic variance account for the level of the plateau we performed a \emph{noiseless} analysis with 6 priors represented by the orange dots in the figure. The cosmic-variance limit reaches $\sigma(\beta_b)\approx0.026\,\deg$. This seems to indicate that for the sky coverage and noise levels of SO SATs the noise accounts for $\sim 40 \%$ of the plateau's amplitude. To improve on the level of the plateau one then needs either to lower the noise or to increase $f_{\rm sky}$.
However with a larger sky survey the spatial variability of foreground SEDs will potentially become a bigger issue for the component separation and might bias the estimation of cosmological parameters.

\subsection{Biased Priors}
\label{sub:bias}
Up to now we have only considered cases where the priors, when averaged over an ensemble of calibration procedures, are centred on the true values of miscalibration angles. In this section, we explore how the framework performs with biased priors. These could for instance arise due to a systematic effect that would not vanish even after the averaging over an ensemble of the calibration procedures as assumed previously. In particular we would like to get some insight on how different biases at different frequency channels would typically impact our conclusions. For this exploratory work we use a simple foreground ``d0s0'' with cosmological parameters set at $r=0$ and $\beta_b= 0^\circ$. All the priors used in this section have a precision of $\sigma_{\alpha_i} = 1^\circ$. We then draw a random set of biases for each frequency channel. They are drawn from a Gaussian distribution centred at $0^\circ$ with a standard deviation of $1^\circ$ following the prior precision.
Hereafter, we explore three cases. First, with only one prior on the 93 GHz channel, which is biased by $0.13^\circ$.
A second case with priors on all channels, where they are all biased by $0.13^\circ$. And the third case with priors on all channels, and all biased by different random amounts. Table~\ref{tab75:input_bias} summarises all these cases.

\begin{table}[]
    \centering
\begin{tabular}{c|c|c|c}
\makecell{Input \\angle [$^\circ$]} & \makecell{Input\\1 prior [$^\circ$]} & \makecell{Input 6 priors,\\equal biases [$^\circ$]} & \makecell{Input 6 priors,\\different biases [$^\circ$]} \\
\cline{1-4}
$\alpha_{27}=$ 1.00 &              & 1.13 ($ 0.13$) & 1.28 ($ 0.28$) \\
$\alpha_{39}=$ 1.67 &              & 1.80 ($ 0.13$) &  0.88 ($ -0.78$) \\
$\alpha_{93}=$ 2.33 &  2.46 ($0.13$) & 2.46 ($ 0.13$) &  2.46 ($ 0.13$) \\
$\alpha_{145}=$ 3.00 &             & 3.13 ($ 0.13$) & 3.30 ($ 0.30$) \\
$\alpha_{225}=$  3.67 &           &  3.80 ($ 0.13$) &  5.12 ($ 1.46$) \\
$\alpha_{280}=$  4.33 &           & 4.46 ($ 0.13$) &   4.43 ($ 0.09$)\\
\cline{1-4}
\end{tabular}
    \caption{The 3 different cases of input biased prior centres used for each frequency channel. The numbers in parentheses correspond to the value of the bias in degrees.
    }
    \label{tab75:input_bias}
\end{table}
The results of the generalised ensemble averaged spectral likelihood sampling for the three cases are displayed in Table~\ref{tab75:res_bias} which shows the recovered biases on estimated parameters.

In the case of the single prior, the \emph{recovered} miscalibration angles for all the channels have the same overall bias of $\Delta \alpha= 0.13^\circ$, which was imposed on the directly calibrated channel of $93$GHz. This is because all these angles are calibrated \emph{relative} to the channel with a prior. We also find that the spectral indices are estimated correctly as their estimates do not depend on the coordinate choice given that the scaling laws assumed for the $Q$ and $U$ Stokes parameters are the same.
In the cosmological likelihood, this bias affects the estimation of $\beta_b$ leading to a bias of $-0.13^\circ$ (see Table~\ref{tab75:res_bias}). Estimation of $r$  is however not impacted by the bias. This is because we estimate $r$ and $\beta_b$ jointly and assume no EB correlations in the CMB covariance. This allows to separate the B-mode signal due to the birefringence or miscalibration from the primordial signal at minimal loss of precision.
We note that the method is similar to self-calibration, which however is performed on a foreground-cleaned and hopefully miscalibration-corrected signal and includes the extra uncertainty due to the marginalisation over the birefringence angle. As we discussed above the method is robust for $r$ even in the case where miscalibration is not well corrected for --- as it is the case here.
\\
In the second case with 6 priors and a $0.13^\circ$ bias on each of them, the estimation of the miscalibration angles are biased by the same value as seen in Table~\ref{tab75:res_bias}. This is like in the single prior case, however with the uncorrelated part of the statistical uncertainties reduced as discussed earlier. This is then biasing $\beta_b$ as well by $-0.13^\circ$, again with no effect on $r$. Finally, in the case where all biases are different we see in the table that all miscalibration angles estimations are biased with the \emph{same} value of $0.25^\circ$, which corresponds to the average of all bias values from the right most column of Table~\ref{tab75:input_bias}. And thanks to the use of many priors, the statistical uncertainty is reduced as well.

We note that as a single prior is sufficient for us to solve the problem, having multiple priors allows for a number of consistency tests, which in actual data analysis practice could shed some light on underlying (albeit unknown) biases.
Finally, the bias on miscalibration angles is carried to $\beta_b$ leading to a $-0.25^\circ$ bias (see Table~\ref{tab75:res_bias}) and $r$ is again not affected.

\begin{table}[]
    \centering
  \begin{tabular}{l|c|c|c}
      & 1 biased prior & \makecell{6 equally\\biased priors} & \makecell{6 differently\\biased priors} \\
  \cline{1-4}
  $\Delta\alpha_{27}[^\circ]$ &  0.13 &    0.13 &    0.24\\
   $\Delta\alpha_{39}[^\circ]$ &  0.13 &    0.13 &    0.24\\
   $\Delta\alpha_{93}[^\circ]$ &  0.13 &    0.13 &    0.25\\
   $\Delta\alpha_{145}[^\circ]$ &  0.13 &    0.13 &    0.25\\
  $\Delta\alpha_{225}[^\circ]$ &   0.13 &    0.13 &    0.25\\
  $\Delta\alpha_{280}[^\circ]$ &  0.13 &    0.13 &    0.25\\
 $\Delta\beta_d$ &  0.00 &    0.00 &    0.00\\
 $\Delta\beta_s$ &  0.00 &    0.00 &  0.00\\
 \cline{1-4}
 $\Delta r$ &    0.00 &    0.00 &    0.00\\
 ${\Delta\beta_b}[^\circ]$ &  -0.13 &  -0.13 &  -0.25
\end{tabular}
    \caption{Resulting bias on the estimation of instrumental parameters, spectral indices and cosmological parameters in the 3 biased cases studied here.}
    \label{tab75:res_bias}
\end{table}

\section{Conclusions}
\label{sec:conclusions}
We propose and demonstrate on simulations a novel method performing component separation in order to remove the contributions due to galactic foregrounds and simultaneously accounting on polarisation-angle misalignment and allowing for an inclusion of calibration priors. The method generalises the pixel-based parametric component separation method introduced in \cite{stompor2009,stompor2016}.
 The instrumental effects are represented via an instrumental response matrix $\boldsymbol{X}$ incorporated directly in the data model. The calibration priors are included as multiplicative terms to the generalised spectral likelihood and the method propagates statistical and systematic errors due to the data, assumed models, and the priors all the way to cosmological parameters.
We focus specifically on the tensor-to-scalar ratio, $r$, and birefringence angle, $\beta_b$, and we fit simultaneously for these two parameters to the CMB signal as recovered from the data on the initial, component separation step.

We specialise this method to forecast performance of a typical future ground based multi-frequency CMB experiments. For this we employ likelihoods semi-analytically averaged over CMB and noise realisations which permits obtaining statistically meaningful predictions and their uncertainties averaged over the same ensemble.

We use the Simons Observatory Small Aperture Telescopes characteristics as our fiducial experimental setup and assume a single, pixel-independent, miscalibration error for each single frequency map. These angles are parameters of the instrumental response matrix, $\mathbf{X}$, and are fitted for on the component separation step together with the parameters describing the foregrounds.
We consider different foreground models, including those where the assumed foreground model matches the actual foreground signal as used in the simulations, and models where we allow for the mismatch between the two. We then investigate the performance of the method from the perspective of the biases and statistical errors on the cosmological parameters.

We show that the data on its own set strong constraints on relative polarisation angles between different single frequency maps.
Consequently, a single prior on a polarisation angle of one of the single frequency maps is sufficient to allow setting meaningful constraints first on polarisation angles for all frequency channels and foreground spectral indices, and later the cosmological parameters.
Using multiple calibration priors on different single frequency maps is beneficial in terms of the resulting statistical uncertainties of the recovered polarisation angles but also allows for robustness tests of the derived results. In our fiducial study cases we find that for a single prior with precision $\sigma_{\alpha_{93}}=0.1^\circ$ on the polarisation angle of the sky map at $93$ GHz, the polarisation angles for all maps can be derived without any biases and with the precision equal to, for the 93GHz channel, and only slightly worse than, for all other channels, than the assumed prior precision.
We find that there is little impact of the more involved data model employed in this work on the estimation of the foreground indices and our results for $r$, $r=0.0002^{+0.0015}_{-0.0018}$, are in agreement with the SO SAT forecasts with \texttt{FGBuster}~\cite{Ade_2019}, which neglect the polarisation angle misalignment. Allowing for the foreground model mismatch does not affect significantly statistical errors but may lead to a bias in estimated values of $r$. However, in the cases studied here the biases were never larger than $1 \sigma$ statistical uncertainties with the most significant bias on  $r$ found in the case of the spatially varying foreground model ``d1s1'', $r=0.0016^{+0.0016}_{-0.0018}$.
We find that the estimates of $r$ are largely independent of the assumed priors and that we can set meaningful constraints on $r$ even in their absence. The proposed method can therefore be considered as
a self-calibration approach.

Priors are necessary however in order to constrain the birefringence parameter.
The foreground model mismatch does not bias the estimates of $\beta_b$, and for the case of a single prior with the precision of $\sigma_{\alpha} = 0.1^\circ$ we get the uncertainty on $\beta_b$ to be $\simeq 0.1^\circ$. For 6 priors with the same precision this improves to $\simeq 0.07^\circ$. In general, the better the priors, the better the final uncertainty on $\beta_b$, however, the latter saturates once the calibration precision gets sufficiently low and the uncertainty on $\beta_b$ starts being dominated by the signal and noise variance. For the studied instrumental setup this happens for calibration precision of $\sim 0.07^\circ$ for a single and $\sim 0.1^\circ$ for 6 calibration priors.
Overall we conclude that the next generation of the CMB polarisation experiments, aiming at the precision of their angle calibration of $\sim 0.1^\circ$, should be capable of rejecting or confirming the value of $0.35^\circ$ suggested by some recent analyses of the Planck data~\cite{Minami_2020, Diego_Palazuelos_2022} with $\sim 3-5 \sigma$ significance depending on the number of the priors.

We also find the biases on the birefringence angle arise only in the cases when the calibration priors themselves are biased. We show however that this does not affect the estimates of $r$.

The instrumental model assumed here is clearly idealised. Most importantly, it neglects band-passes. These would affect both the actual sky signal but also polarisation angle calibration in the way which will depend on their effective spectral dependence. To first order this will lead to biases on the priors and therefore including such effects is a key to any claim about the detection of birefringence. This could also affect the $r$ constraints but mostly via their impact on the foreground residuals. Other effects which could affect the polarisation angle such as smoothly rotating half-wave plate or sinuous antennas, are also relevant and should be taken into account. We leave such extensions to future work.

Similarly unrealistic is the assumption of a single pixel-independent polarisation angle per a single frequency map. Indeed, the miscalibration angle should be more of a property of a detector or of a focal plane wafer, this will generally lead to a pixel-domain effective polarisation angle on the map-level due to the fact that different wafers/detectors typically observe the sky differently. This formalism is easily adaptable to using as an input maps produced for every wafer or detector, each with a specific polarisation angle. This may however lead to proliferation of the instrumental degrees of freedom in the spectral likelihood problem with potential effects on the precision of derived constraints. More studies are needed to assess whether this can be successfully controlled.

Complex foregrounds do not seem to significantly affect the estimation of polarisation angles however the small sky and frequency coverage of the SATs limits the impact of complex foregrounds with respect to simpler ones. \\

\section*{Acknowledgements}
We thank Clara Verg\`es, Cl\'ement Leloup, Hamza El Bouhargani, Magdy Morshed, Arianna Rizzieri and Simon Biquard for useful discussions. This research used resources of the National Energy Research Scientific Computing Center (NERSC), a U.S. Department of Energy Office of Science User Facility located at Lawrence Berkeley National Laboratory. The authors acknowledge support of the French National Research Agency (Agence National de Recherche) grant, ANR BxB (ANR-17-CE31-0022) and B3DCMB (ANR-17-CE23-0002). This work is also part of a project that has received funding from the European Research Council (ERC) under the European Union’s Horizon 2020 research and innovation program (PI: Josquin Errard, Grant agreement No. 101044073).
Some of the results in this paper have been derived using the \texttt{healpy}, \texttt{numpy} and \texttt{PySM} packages. Some of the figures in this article have been created using \texttt{GetDist}.

\bibliography{biblio}

\end{document}